\documentclass[aps,pre,showpacs,floats,preprint]{revtex4}

\usepackage{amssymb}
\usepackage{graphics}
\usepackage[dvipdfm]{graphicx}
\usepackage{epsfig}
\usepackage{dcolumn}
\usepackage{amsmath}

\begin{document}

\title{Response Function Theory for Many-Body Systems away-from Equilibrium: Conditions of
Ultrafast-Time and Ultrasmall-Space Experimental Resolution}
\author{Cl\'{o}ves G. Rodrigues$^{1}$, \'{A}urea R.
Vasconcellos$^{2}$, J. Galv\~{a}o Ramos$^{2}$, Roberto
Luzzi$^{2}$\footnote{group home page:
www.ifi.unicamp.br/$\sim$aurea}} \affiliation{$^{1}$Departamento de
F\'{\i}sica, Pontif\'{\i}cia Universidade Cat\'{o}lica
de Goi\'{a}s, 74605-010 Goi\^{a}nia, Goi\'{a}s, Brazil\\
$^{2}$Condensed Matter Physics Department, Institute of Physics
\textquotedblleft Gleb Wataghin'',\\
University of Campinas-Unicamp, 13083-859 Campinas, SP, Brazil}
\date{\today }

\begin{abstract}
A Response Function Theory and Scattering Theory applicable to the
study of physical properties of systems driven arbitrarily away from
equilibrium, specialized for dealing with ultrafast processes and in
conditions of space resolution (including nanometric scale), are
presented. The derivation is done in the framework of a Gibbs-style
Nonequilibrium Statistical Ensemble Formalism. It is shown the
connection of the observable properties with time and
space-dependent correlation functions out of equilibrium. A
generalized fluctuation-dissipation theorem, which relates these
correlation functions with generalized susceptibilities is derived.
It is also presented the method, useful for calculations, of
nonequilibrium-thermodynamic Green functions. A couple of
illustration with application of the formalism, consisting of the
study of optical responses in ultrafast laser spectroscopy and Raman
Scattering of electrons in III-N semiconductors (of "blue diodes")
driven away from equilibrium by action of electric fields of
moderate to high intensities, are described.
\end{abstract}

\pacs{01.75.+m; 01.78.+p; 05.70.Ln; 05.90.+m; 05.60.Gg; 89.75.-k;
81.05.Ea}
\maketitle

\section{Introduction}

The renowned Ryogo Kubo once stated that "statistical mechanics has
been considered a theoretical endeavor. However, statistical
mechanics exists for the sake of the real world, not for fictions.
Further progress can only be hoped by closed cooperation with
experiment" [l]. This is nowadays particularly relevant because the
notable development of all the modern technology, fundamental for
the progress and well being of the world society, poses a great deal
of stress in the realm of basic Physics, more precisely on
Thermo-Statistics. Thus, on the one hand, we do face situations in
electronics and optoelectronics involving physical-chemical systems
far-removed-from equilibrium, where ultrafast (pico- and
femto-second scale) and non-linear processes are present. Further,
we need to be aware of the rapid unfolding of nano-technologies and
use of low-dimensional systems (e.g., nanometric quantum wells and
quantum dots in semiconductors heterostructures) [2]. All together
this demands having an access to a statistical mechanics being
efficient to deal with such requirements. On the other hand, one
needs to face the study of soft matter and fluids with complex
structures (usually of the average self-affine fractal-like type)
[3]. This is relevant for technological improvement in industries
like, for example, that of polymers, petroleum, cosmetics, food,
electronics and photonics (conducting polymers and glasses), in
medical engineering, etc. Moreover, in both type of situations above
mentioned there often appear difficulties of description and
objectivity (existence of so-called "hidden constraints"), which
impair the proper application of the conventional ensemble approach
used in the general, logically and physically sound, and well
established Boltzmann-Gibbs statistics. A tentative to partially
overcome such difficulties consists into resorting to
non-conventional approaches [4-7].

Since, as noticed, a most relevant objective of any nonequilibrium
statistical theory is to provide a comprehension of the underlying
physics related to the relaxation phenomena that can be evidenced in
experiments, it needs be coupled with a response function theory.
This is the subject of this paper, where we specifically resort to
the use of a Non-Equilibrium Statistical Ensemble Formalism (NESEF
for short) [8-11]).

It can be noticed that nowadays two approaches appear to be the most
favorable for providing very satisfactory methods to deal with
systems within an ample scope of nonequilibrium conditions. They
are, on the one hand, Numerical Simulation Methods [12], or
Computational Physics. In particular, to it belongs Non-Equilibrium
Molecular-dynamics NMD [13], a computational method created for
modeling physical systems at the microscopic level, being a good
technique to study the molecular behavior of several physical
processes. On the other hand, we do have the kinetic theory based on
the far-reaching generalization of Gibbs' ensemble formalism, the
NESEF [11,14].

The present structure of the formalism consists in an extension and
generalization of earlier pioneering approaches, among which we can
pinpoint the works of Kirkwood [15], Green [16], Mori-Oppenheim-Ross
[17], Mori [18] and Zwanzig [19]. NESEF has been approached from
different points of view: some are based on heuristic arguments,
others on projection-operator techniques (the former following
Kirkwood and Green and the latter following Zwanzig and Mori).

The formalism has been systematized and largely improved by the
Russian School of statistical physics, which can be considered to
have been initiated by the renowned Nicolai Nicolaievich Bogoliubov
(e.g., see ref. [20]) and we may also name Nicolai Sergeievich
Krylov [21], and more recent1y mainly through the relevant
contributions of Dimitrii Zubarev [8,9], Sergei Peletminskii [22],
and others. We present in Refs. [11] a systematization, as well as
generalizations and conceptual discussions, of the matter.

It may be noticed that these different approaches to NESEF can be
brought together under a unique variational principle. This has been
originally done by Zubarev and Kalashnikov [23] and later on
reconsidered in Refs. [9,11]. It consists on the maximization, in
the context of information Theory, of Gibbs statistical entropy
(that is, the average of minus the logarithm of the statistical
distribution function [24,25], which in Communication Theory is
Shannon informational entropy [26,27], subjected to certain
constraints and including non-locality in space, retro-effects, and
irreversibility on the macroscopic level.

Concerning Response Function Theory, the usual theory to calculate
linear responses to mechanical perturbations (e.g. [28-33]) is based
on expansions in terms of correlation functions in equilibrium. As
initial condition is taken that of equilibrium with a thermal
reservoir, and next it is studied the evolution of the system as if
it were isolated from all external influences except the driving
field. Let us now consider the situation when a mechanical
perturbation is applied on an already far-from-equilibrium system,
in which are unfolding irreversible processes which are describable
in terms of equations of evolution for a basic set of macrovariables
in the non-equilibrium thermodynamic space of states. Since NESEF
provides a seemingly powerful method to obtain a description of the
macrostate of such systems, it is appealing to derive a response
function theory based on correlation functions in the unperturbed
nonequilibrium state of the system. Schemes of this type have been
proposed [31-33], and next we systematize and extend this treatment,
in such a way to allow the treatment of experiments involving
time-resolution (including the ultrafast time scale of pico- and
femto-seconds), and space resolution (including those in the
emerging nano-science and technology).

It is shown the connection of the observable properties with
correlation functions out of equilibrium; a generalized
fluctuation-dissipation theorem -- relating correlation functions
and generalized susceptibilities -- is derived, and the method,
useful for calculations, of nonequilibrium-thermodynamic Green
functions is presented. This is done in Sections II, III and IV, and
in Section V we present a scattering theory, in the same conditions,
namely, including time and space resolution, for systems far-from
equilibrium. The connection of the scattering theory with response
function theory follows from application of the nonequilibrium
fluctuation-dissipation theorem.

Finally, in Section VI we present a couple of illustrations showing
the working of the theory in the study of two kind of experiments,
namely optical responses in ultrafast laser spectroscopy of polar
semiconductors, and Raman scattering of electrons in doped III-N
semiconductors ("blue diodes") in the presence of electric fields
with moderate to high intensities. In the latter case the
nonequilibrium fluctuation-dissipation theorem allows to connect the
Raman spectrum with nonlinear transport properties in these
materials (nonlinear and time-dependent conductivity and diffusion
coefficient, and a generalized - nonlinear and time-dependent
Einstein-relation).


\section{Response Function Theory for Far-From-Equilibrium Systems}

We consider an open many-body system out of equilibrium, which is in
contact with a set of reservoirs and under the action of pumping
sources. We are essentially presenting the most general experiment
one can think of, namely a sample (the open system of interest
composed of very-many degrees of freedom) subjected to given
experimental conditions, as it is diagrammatically described in Fig.
1.

%
\begin{figure}[h]
\center
\includegraphics[width=10cm]{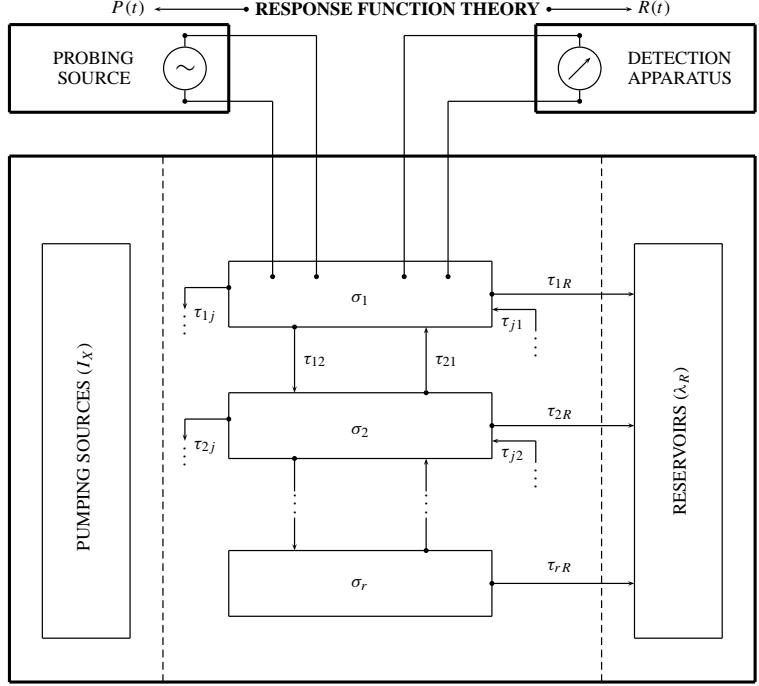}
\caption{Diagrammatic description of a typical pump-probe experiment in an
open dissipative system.}
\end{figure}

In Fig. 1, the sample is composed of a number of subsystems, $\sigma_j$, (or
better to say subdegrees of freedom, for example, in solid state matter
those associated to electrons, lattice vibrations, excitons, impurity
states, collective excitations as plasmons, magnons, etc., hybrid
excitations as polarons, polaritons, plasmaritons and so on). They interact
among themselves via interaction potentials producing exchange at certain
rates, $\tau_{ij}$, of energy and momentum. Pumping sources act on the
different subsystems of the sample -- via particular types of fields,
electric, magnetic, electromagnetic, etc. -- well characterized when setting
up the experiment, and there follows relaxation of the energy in excess of
equilibrium the system is receiving to the external reservoirs, $\tau_{jR}$.
Finally, the experiment is performed coupling an external probing source,
characterized in the figure by $P(t)$, with one or more subsystems of the
sample, and some kind of response, say $R(t)$, is detected by a measuring
apparatus (e.g. ammeter, spectrometer, etc.).

It needs be understood that the pumping sources exert their influence on the
given open system through the fields they generate, say, magnetic, electric,
electromagnetic as produced for example from a laser machine, and so on, or,
eventually, in scattering experiments is the interaction potential with the
particles of an incoming beam.

Let the Hamiltonian be
%
\begin{equation}
\Hat{H}(t)=\Hat{H}_{\sigma }+\Hat{\mathcal{V}}(t)\;,
\end{equation}
that is, the Hamiltonian $\Hat{H}_{\sigma }$ of the system in the
presence of the fields of the pumping sources which drive it away
from equilibrium, plus the interactions with the reservoirs, and
with the field(s), $\Hat{\mathcal{V}}(t)$, created by the external
perturbing apparatus. We take for the latter the form
%
\begin{equation}
\Hat{\mathcal{V}}(t)=-\int d^{3}r^{\prime }\,\mathcal{F}(\mathbf{r}^{\prime
},t)\Hat{A}(\mathbf{r}^{\prime })\;,
\end{equation}
where $\mathcal{F}$ is the expression for the perturbing force
($\mathcal{F}=-\delta \Hat{\mathcal{V}}/\delta
\Hat{A}(\mathbf{r}^{\prime })$, where $\delta $ is a functional
derivative) and $\Hat{A}(\mathbf{r}^{\prime })$ an observable of the
system to which it is coupled. We recall that the system is in
contact with ideal reservoirs, and the statistical operator is taken
as a product of the one of the system  $\varrho _{\varepsilon }(t)$
times the stationary canonical distribution of the reservoirs
$\varrho_{R}$, which we write $\mathcal{R}_{\varepsilon }(t)=\varrho
_{\varepsilon }(t)\times \varrho _{R}$ [11] (see Appendix A).
Moreover, $\Hat{H}_{\sigma
}=\Hat{H}_{0}+\Hat{H}_{1}+\Hat{W}=\Hat{H}_{0}+\Hat{H}^{\prime }$,
introducing $\Hat{H}^{\prime }=\Hat{H}_{1}+\Hat{W}$, where $H_{1}$
accounts for all the interactions in the system and $W$ for the
interaction with the surroundings.

Schr\"odinger equation for this system, i.e.
%
\begin{equation}
i\hbar\frac{\partial}{\partial t}|\psi(t)\rangle=\Hat{H}(t)|\psi(t)\rangle\;,
\end{equation}
with the initial condition $|\psi(t_i)\rangle$ at time $t_i$ when the
perturbation is switched on, has the formal solution
%
\begin{equation}
|\psi(t)\rangle = U(t,t_i)|\psi(t_i)\rangle\;,
\end{equation}
where $U$ is the evolution operator satisfying that
%
\begin{equation}
i\hbar\frac{\partial}{\partial t} U(t,t_i)=\Hat{H}(t)U(t,t_i)\;,
\end{equation}
with $U(t_i,t_i)=\hat{1}$ (the unit operator), and we recall that it is a
unitary operator, that is $U^{\dagger}U=\hat{1}$.

Let us now introduce the interaction representation, writing
%
\begin{equation}
U(t,t_i)=U_{\sigma}(t,t_i)U^{\prime}(t,t_i)\;,
\end{equation}
where
%
\begin{equation}
U_{\sigma}(t,t_i)=\exp \Biggl\{
\frac{1}{i\hbar}(t-t_i)\Hat{H}_{\sigma} \Biggr\}\;,
\end{equation}
which is the evolution operator in the absence of the perturbation
potential $\Hat{\mathcal{V}}$; this means that we are separating the
internal dynamics of the system from the dynamical effects of the
perturbation which are accounted for by $U^{\prime}$. Introducing
Eq. (6) in Eq. (5) and using Eq. (7), it follows that $U^{\prime}$
satisfies the equation
%
\begin{equation}
i\hbar\frac{\partial}{\partial
t}U^{\prime}(t,t_i)=\tilde{\mathcal{V}}(t)U^{\prime}(t,t_i)\;,
\end{equation}
with $U^{\prime}(t_i,t_i)=\hat{1}$, and
%
\begin{equation}
\tilde{\mathcal{V}}(t)=U_{\sigma}^{\dagger}(t,t_i)\Hat{\mathcal{V}}(t)
U_{\sigma}(t,t_i)\;.
\end{equation}

Equation (8) can be transformed in an equivalent integral equation,
namely
%
\begin{equation}
U^{\prime}(t,t_i)=\hat{1}+\frac{1}{i\hbar}\int_{t_i}^t
dt^{\prime}\,\tilde{
\mathcal{V}}(t^{\prime})U^{\prime}(t^{\prime},t_i)\;,
\end{equation}
which admits the iterated solution
%
\begin{eqnarray}
U^{\prime }(t,t_{i}) &=&\hat{1}+\frac{1}{i\hbar
}\int_{t_{i}}^{t}dt^{\prime
}\,\tilde{\mathcal{V}}(t^{\prime })+\frac{1}{(i\hbar )^{2}}%
\int_{t_{i}}^{t}dt^{\prime }\int_{t_{i}}^{t^{\prime }}dt^{\prime
\prime }\, \tilde{\mathcal{V}}(t^{\prime
})\tilde{\mathcal{V}}(t^{\prime \prime
})+\cdots   \notag \\
&=&\hat{1}+\sum_{n=1}^{\infty }\frac{1}{(i\hbar
)^{n}}\int_{t_{i}}^{t}dt_{1} \cdots
\int_{t_{i}}^{t_{n-1}}dt_{n}\,\tilde{\mathcal{V}}(t_{1})\cdots
\tilde{ \mathcal{V}}(t_{n})\;.
\end{eqnarray}

The quantum-mechanical expected value at time $t$ of the observable
$\Hat{A}(\mathbf{r})$, to which the external field is coupled, is
then
%
\begin{eqnarray}
a(\mathbf{r},t) &=&\langle \psi (t)|\Hat{A}(\mathbf{r})|\psi
(t)\rangle
\notag \\
&=&\langle \psi (t_{i})|{U^{\prime }}^{\dagger }(t,t_{i})U_{\sigma
}^{\dagger }(t,t_{i})\Hat{A}(\mathbf{r})U_{\sigma
}(t,t_{i})U^{\prime
}(t,t_{i})|\psi (t_{i})\rangle   \notag \\
&=&\langle \psi (t_{i})|{U^{\prime }}^{\dagger
}(t,t_{i})\tilde{A}(\mathbf{r} ,t)U^{\prime }(t,t_{i})|\psi
(t_{i})\rangle \;,
\end{eqnarray}
where
%
\begin{equation}
\tilde{A}(\mathbf{r},t)=U_{\sigma }^{\dagger
}(t,t_{i})\Hat{A}(\mathbf{r})U_{\sigma }(t,t_{i})\;.
\end{equation}

According to Eq. (11) we have that
%
\begin{eqnarray}
a(\mathbf{r},t) &=&\langle \psi (t_{i})|(\hat{1}-\frac{1}{i\hbar }
\int_{t_{i}}^{t}dt^{\prime }\,\tilde{\mathcal{V}}(t^{\prime
})+\cdots )
\tilde{A}(\mathbf{r},t)\times   \notag \\
&&(\hat{1}+\frac{1}{i\hbar }\int_{t_{i}}^{t}dt^{\prime
}\,\tilde{\mathcal{V}} (t^{\prime })+\cdots )|\psi (t_{i})\rangle
\;,
\end{eqnarray}
where we have used that $\tilde{\mathcal{V}}$ is Hermitian.

Considering a weak perturbation -- characterized by
$\Hat{\mathcal{V}}(t)$ -- imposed on the initially (at time $t_{i}$)
far-from-equilibrium system, we truncate the series of terms in Eq.
(14) in first order in $\tilde{\mathcal{V}}$, that is, we consider
from now on a \textit{linear response theory,} to obtain that
%
\begin{equation}
a(\mathbf{r},t)=a(\mathbf{r},t_{i})+\frac{1}{i\hbar
}\int_{t_{i}}^{t}dt^{ \prime }\,\langle \psi
(t_{i})|[\tilde{A}(\mathbf{r},t),\tilde{\mathcal{V}}(t^{\prime
})]|\psi (t_{i})\rangle \;,
\end{equation}
where
%
\begin{equation}
a(\mathbf{r},t_{i})=\langle \psi (t_{i})|\Hat{A}(\mathbf{r})|\psi
(t_{i})\rangle
\end{equation}
is the expected value of the observable at time $t_{i}$ prior to the
application of the perturbation.

Using Eq. (2) we have that
%
\begin{equation}
\Delta a(\mathbf{r},t)=-\frac{1}{i\hbar
}\int_{t_{i}}^{t}dt^{\prime }\int d^{3}r^{\prime }\,\langle \psi
(t_{i})|[\tilde{A}(\mathbf{r},t),\tilde{A}(\mathbf{r}^{\prime
},t^{\prime })]\mathcal{F}(\mathbf{r}^{\prime },t^{\prime })|\psi
(t_{i})\rangle \;,
\end{equation}
where $\Delta a(\mathbf{r},t)=a(\mathbf{r},t)-a(\mathbf{r},t_{i})$ is the
departure of the observable $\Hat{A}$ from its value at the initial time
when under the action of the perturbing potential.

Introducing the statistical operator for the pure (quantum mechanical)
state, namely
%
\begin{equation}
\mathcal{P}(t_{i})=|\psi (t_{i})\rangle \langle \psi (t_{i})|\;,
\end{equation}
which, we recall, is a projection operator over the vector state
$|\psi (t_{i})\rangle $, we can rewrite Eq. (17) as
%
\begin{equation}
\Delta a(\mathbf{r},t)=-\frac{1}{i\hbar }\int_{-\infty }^{t}dt^{\prime }\int
d^{3}r^{\prime }\,Tr\{[\tilde{A}(\mathbf{r},t),\tilde{A}(\mathbf{r}^{\prime
},t^{\prime })]\mathcal{P}(t_{i})\}\mathcal{F}(\mathbf{r}^{\prime
},t^{\prime })\;,
\end{equation}
and where we have considered adiabatic application of the perturbation taken
as \hbox{$t_i\rightarrow -\infty$}.

Next step is going over the macroscopic state taking the average
over the nonequilibrium ensemble of pure states, compatible with
the macroscopic conditions of preparation of the sample. In the
usual way, if we call $\mathcal{P}_{n}(t_{i})$ the statistical
operator for the pure state in, say, the $n$-th replica, and
$p_{n}$ the probability of such replica in the corresponding Gibbs
ensemble, the statistical average over the ensemble of mixed
states of the system, and the average over the states of the
reservoir (system and reservoir are coupled via the interaction
$\Hat{W}$) is
%
\begin{eqnarray}
\Delta \langle \Hat{A}(\mathbf{r})|t\rangle  &=&-\frac{1}{i\hbar }
\int_{-\infty }^{t}\!\!dt^{\prime }\int \!d^{3}r^{\prime
}\sum_{n}p_{n}\times   \notag \\
&&Tr\bigl\{\lbrack
\tilde{A}(\mathbf{r},t),\tilde{A}(\mathbf{r}^{\prime },t^{\prime
})]\mathcal{P}_{n}(t_{i})\times \varrho _{R}\bigr\}\mathcal{F}(
\mathbf{r}^{\prime },t^{\prime })  \notag \\
&=&-\frac{1}{i\hbar }\int_{-\infty }^{t}dt^{\prime }\int
d^{3}r^{\prime
}\,Tr\{[\tilde{A}(\mathbf{r},t),\tilde{A}(\mathbf{r}^{\prime
},t^{\prime })]\varrho _{\varepsilon }(t_{i})\times \varrho
_{R}\}\mathcal{F}(\mathbf{r}^{\prime },t^{\prime })\;,
\end{eqnarray}
with
%
\begin{equation}
\varrho _{\epsilon }(t_{i})=\sum_{n}p_{n}\mathcal{P}_{n}(t_{i})=
\sum_{n}p_{n}|\psi _{n}(t_{i})\rangle \langle \psi _{n}(t_{i})|\;,
\end{equation}
and we have introduced
%
\begin{equation}
\Delta \langle \Hat{A}(\mathbf{r})|t\rangle
=Tr\{\Hat{A}(\mathbf{r})\mathcal{R}_{\varepsilon
}(t)\}-Tr\{\Hat{A}(\mathbf{r})\mathcal{R}_{\varepsilon }(t_{i})\}\;,
\end{equation}
where, we recall, $\mathcal{R}_{\varepsilon }(t)=\varrho _{\varepsilon
}(t)\times \varrho _{R}$.

But Eq. (20) can be rewritten as
%
\begin{equation}
\Delta \langle \Hat{A}(\mathbf{r})|t\rangle =-\frac{1}{i\hbar
}\int_{-\infty }^{t}dt^{\prime }\int
d^{3}r,Tr\bigl\{[\tilde{A}(\mathbf{r}),\tilde{A}(
\mathbf{r}^{\prime },t^{\prime }-t)]\mathcal{R}_{\varepsilon
}(t)\bigr\}\mathcal{F}(\mathbf{r}^{\prime },t^{\prime })\;,
\end{equation}
since
%
\begin{multline}
Tr\{[U_{\sigma }^{\dagger }(t,t_{i})\Hat{A}(\mathbf{r})U_{\sigma
}(t,t_{i}),U_{\sigma }^{\dagger }(t^{\prime
},t_{i})\Hat{A}(\mathbf{r}^{\prime })U_{\sigma }(t^{\prime
},t_{i})]\mathcal{R}_{\varepsilon
}(t_{i})\}= \\
Tr\{\tilde{A}(\mathbf{r}),\tilde{A}(\mathbf{r}^{\prime },t^{\prime
}-t)\mathcal{R}_{\varepsilon }(t)\}\;,
\end{multline}
where we have used the invariance of the trace operation by cyclical
permutations and the group properties of operators $U$. We stress
that in nonequilibrium conditions there is no invariance for
translation in time as a result of the time-dependence of the
statistical operator which describes the irreversible evolution of
the macroscopic state; the operator $A(t^{\prime }-t)$ in Eq. (24)
shows time-translational invariance because such dependence on time
arises out of the microdynamical evolution governed by Hamiltonian
$H_{\sigma }$.

Moreover, introducing the definitions
%
\begin{equation}
\chi ^{\prime \prime }(\mathbf{r},\mathbf{r}^{\prime };t^{\prime
}-t|t)= \frac{1}{2\hbar
}Tr\{[\Hat{A}(\mathbf{r}),\tilde{A}(\mathbf{r}^{\prime },t^{\prime
}-t)]\mathcal{R}_{\varepsilon }(t)\}\;,
\end{equation}
%
\begin{equation}
\chi (\mathbf{r},\mathbf{r}^{\prime };t-t^{\prime }|t)=2i\theta
(t-t^{\prime })\chi ^{\prime \prime }(\mathbf{r},\mathbf{r}^{\prime
};t^{\prime }-t|t)\;,
\end{equation}
with $\theta $ being Heaviside's step function, we have that Eq.
(20) becomes
%
\begin{equation}
\Delta \langle \Hat{A}(\mathbf{r})|t\rangle =\int_{-\infty }^{\infty
}dt^{\prime }\int d^{3}r^{\prime }\,\chi (\mathbf{r},\mathbf{r}^{\prime
};t-t^{\prime }|t)\mathcal{F}(\mathbf{r}^{\prime },t^{\prime })\;.
\end{equation}

Taking into account the integral representation of Heaviside step
function, namely
%
\begin{equation}
\theta (\tau )=\lim_{s\rightarrow +\!0}\;i\int_{-\infty }^{\infty
}\frac{d\omega }{2\pi }\frac{e^{-i\omega \tau }}{\omega +is}\;,
\end{equation}
we find, after some calculus, that the Fourier transform in time
$\tau =t-t^{\prime }$ of the nonequilibrium generalized
susceptibility of Eq. (26) is given by
%
\begin{equation}
\chi (\mathbf{r},\mathbf{r}^{\prime };\omega |t)=\int_{-\infty
}^{\infty } \frac{d\omega ^{\prime }}{\pi }\frac{\chi ^{\prime
\prime }(\mathbf{r}, \mathbf{r}^{\prime };\omega ^{\prime
}|t)}{\omega ^{\prime }-\omega -is}\;,
\end{equation}
where, we recall, $s\rightarrow +\!0$, and $\chi ^{\prime \prime
}(\mathbf{r},\mathbf{r}^{\prime };\omega ^{\prime }|t)$ is the
Fourier transform at frequency $\omega $ of the $\tau $-dependent
$\chi ^{\prime \prime }$ of Eq. (25), namely
%
\begin{equation}
\chi ^{\prime \prime }(\mathbf{r},\mathbf{r}^{\prime };\omega
|t)=\int_{-\infty }^{\infty }d\tau \,\chi ^{\prime \prime
}(\mathbf{r}, \mathbf{r}^{\prime };-\tau |t)e^{i\omega \tau }\;.
\end{equation}

Using the fact that
%
\begin{equation}
\lim_{s\rightarrow +\!0}\frac{1}{x\pm is}=pv\frac{1}{x}\mp i\pi \delta (x)\;,
\end{equation}
which are the so-called advanced and retarded Heisenberg delta
functions and $pv$ stands for principal value, Eq. (29) becomes
%
\begin{eqnarray}
\chi (\mathbf{r},\mathbf{r}^{\prime };\omega |t) &=&Re\{\chi
(\mathbf{r}, \mathbf{r}^{\prime };\omega |t)\}+iIm\{\chi
(\mathbf{r},\mathbf{r}^{\prime
};\omega |t)\}  \notag \\
&=&pv\int_{-\infty }^{\infty }\frac{d\omega ^{\prime }}{\pi
}\frac{\chi ^{\prime \prime }(\mathbf{r},\mathbf{r}^{\prime };\omega
^{\prime }|t)}{ \omega ^{\prime }-\omega }+i\chi ^{\prime \prime
}(\mathbf{r},\mathbf{r} ^{\prime };\omega |t)\;,
\end{eqnarray}
where $Re$ and $Im$ stand for real and imaginary parts respectively, and
then
%
\begin{equation}
Im\chi (\mathbf{r},\mathbf{r}^{\prime };\omega |t)=\chi ^{\prime
\prime }( \mathbf{r},\mathbf{r}^{\prime };\omega |t)\;,
\end{equation}
%
\begin{equation}
Re\chi (\mathbf{r},\mathbf{r}^{\prime };\omega |t)=pv\int_{-\infty
}^{\infty }\frac{d\omega ^{\prime }}{\pi }\frac{Im\chi
(\mathbf{r},\mathbf{r}^{\prime };\omega ^{\prime }|t)}{\omega
^{\prime }-\omega }\;,
\end{equation}
after recalling that $\chi ^{\prime \prime
}(\mathbf{r},\mathbf{r}^{\prime };\omega |t)$ is a real quantity as
shown in Eq. (51).

This Eq. (34) is one of the so-called \textit{generalized
Kramer-Kr\"{o}nig relations,} with the other being
%
\begin{equation}
Im\chi (\mathbf{r},\mathbf{r}^{\prime };\omega
|t)=-pv\int_{-\infty }^{\infty }\frac{d\omega ^{\prime }}{\pi
}\frac{Re\chi (\mathbf{r},\mathbf{r}^{\prime };\omega ^{\prime
}|t)}{\omega ^{\prime }-\omega }\;,
\end{equation}
obtained from Eq. (34) once it is used the operational relation
%
\begin{equation}
pv\int_{-\infty }^{\infty }\frac{d\omega ^{\prime \prime }}{\pi }(\omega
-\omega ^{\prime \prime })^{-1}(\omega ^{\prime \prime }-\omega ^{\prime
})^{-1}=-\pi \delta (\omega -\omega ^{\prime })\;.
\end{equation}

We recall that Kramer-Kr\"{o}nig relations are a consequence of the
principle of causality, and involving the fact that $\chi $, once
extended to the complex $z$-plane ($z=\omega +iy$), i.e. $\chi
(\mathbf{r},\mathbf{r}^{\prime };z|t)$, has poles in the lower
$z$-plane and it is regular in the upper $z$-plane. Furthermore we
notice that $Re\chi $ is an even function of $\omega $ while $Im\chi
$ is an odd one. Thus, $\chi (\mathbf{r},\mathbf{r}^{\prime };\omega
|t)$ has the same properties as the equilibrium one presented in Eq.
(52) below, as it should.

Moreover, $Im\chi (\mathbf{r},\mathbf{r}^{\prime };\omega |t)$ is related to
the power absorption by the system. First, we notice that the external force
applied on the system can be Fourier analyzed in time, obtaining a linear
superposition of Fourier components, namely
%
\begin{equation}
\mathcal{F}(\mathbf{r}^{\prime },t^{\prime })=\int d\omega
\frac{1}{2}\mathcal{F}(\mathbf{r}^{\prime },\omega )(e^{i\omega
t^{\prime }}+e^{-i\omega t^{\prime }})\;,
\end{equation}
so it is a real quantity, and let us calculate the average over
time $t^{\prime }$ (and then $\tau $) of the quantity which
represents the power absorbed by the system, namely the average
over a time interval $\Delta t$ (typically the experimental
resolution time) given by
%
\begin{eqnarray}
W(t) &=&\frac{1}{\Delta t}\int_{t}^{t+\Delta t}dt^{\prime }\,\frac{
dE(t^{\prime })}{dt^{\prime }}=\frac{1}{\Delta t}\int_{t}^{t+\Delta
t}dt^{\prime }\,\frac{d}{dt^{\prime }}Tr\{\Hat{H}(t^{\prime
})\mathcal{R}
_{\varepsilon }(t^{\prime })\}  \notag \\
&=&\frac{1}{\Delta t}\int_{t}^{t+\Delta t}dt^{\prime
}Tr\Biggl\{\frac{
\partial \Hat{H}(t^{\prime })}{\partial t^{\prime }}\mathcal{R}_{\varepsilon
}(t^{\prime })+\Hat{H}(t^{\prime })\frac{\partial }{\partial
t^{\prime }} \mathcal{R}_{\varepsilon }(t^{\prime })\Biggr\}\;,
\end{eqnarray}

But, we do have that
%
\begin{equation}
Tr\Bigl\{\Hat{H}(t^{\prime })\frac{\partial }{\partial t^{\prime
}}\mathcal{R }_{\varepsilon }(t^{\prime })\Bigr\}=\frac{1}{i\hbar
}Tr\bigl\{[\Hat{H} (t^{\prime }),\Hat{H}(t^{\prime
})]\mathcal{R}_{\varepsilon }(t^{\prime }) \bigr\}=0\;,
\end{equation}
where we have used that $\mathcal{R}_{\varepsilon }$ satisfies Liouville
equation. Hence
%
\begin{eqnarray}
W(t) &=&\frac{1}{\Delta t}\int_{t}^{t+\Delta t}dt^{\prime
}\,Tr\Bigl\{\frac{
\partial \Hat{H}(t^{\prime })}{\partial t^{\prime }}\mathcal{R}_{\varepsilon
}(t^{\prime })\Bigr\}  \notag \\
&=&\frac{1}{\Delta t}\int_{t}^{t+\Delta t}dt^{\prime
}\,Tr\Bigl\{\frac{
\partial \Hat{\mathcal{V}}(t^{\prime })}{\partial t^{\prime }}\mathcal{R}
_{\varepsilon }(t^{\prime })\Bigr\}  \notag \\
&=&-\frac{1}{\Delta t}\int_{t}^{t+\Delta t}dt^{\prime }\int
d^{3}r^{\prime }Tr\{\Hat{A}(\mathbf{r}^{\prime
})\mathcal{R}_{\varepsilon }(t^{\prime })\} \frac{\partial
\mathcal{F}(\mathbf{r}^{\prime },t^{\prime })}{\partial t^{\prime
}}\;,
\end{eqnarray}
with $\Hat{\mathcal{V}}$ given in Eq. (2). Taking into account that
%
\begin{eqnarray}
Tr\{\Hat{A}(\mathbf{r}^{\prime })\mathcal{R}_{\varepsilon }(t)\}
&=&Tr\{\Hat{ A}(\mathbf{r}^{\prime })U(t^{\prime
},t_{i})\mathcal{R}_{\varepsilon
}(t_{i})U^{\dagger }(t^{\prime },t_{i})\}  \notag \\
&=&Tr\{U^{\dagger }(t^{\prime },t_{i})\Hat{A}(\mathbf{r}^{\prime
})U(t^{\prime },t_{i})\mathcal{R}_{\varepsilon }(t_{i})\}  \notag \\
&=&Tr\{{U^{\prime }}^{\dagger }(t^{\prime
},t_{i})\tilde{A}(\mathbf{r} ^{\prime },t^{\prime })U^{\prime
}(t^{\prime },t_{i})\mathcal{R}_{\varepsilon }(t_{i})\}\;,
\end{eqnarray}
where we used Eqs. (6) and (13), and next, resorting to Eq. (11) in
first order (linear response) and to Eq. (20), we find that
%
\begin{eqnarray}
W(t) &=&\frac{1}{i\hbar \Delta t}\int_{t}^{t+\Delta t}dt^{\prime
}\int d^{3}r^{\prime }\int_{-\infty }^{t^{\prime }}dt^{\prime \prime
}\int d^{3}r^{\prime \prime }\,Tr\{[\tilde{A}(\mathbf{r}^{\prime
},t^{\prime }),
\tilde{A}(\mathbf{r}^{\prime \prime },t^{\prime \prime })]\times   \notag \\
&&\mathcal{R}_{\varepsilon }(t_{i})\}\mathcal{F}(\mathbf{r}^{\prime
\prime },t^{\prime \prime })\frac{\partial
\mathcal{F}(\mathbf{r}^{\prime
},t^{\prime })}{\partial t^{\prime }}  \notag \\
&=&-\frac{1}{\Delta t}\!\int_{t}^{t+\Delta t}\!\!dt^{\prime }\!\int
\!d^{3}r^{\prime }\,\Delta A(\mathbf{r}^{\prime },t^{\prime
})\frac{\partial \mathcal{F}(\mathbf{r}^{\prime },t^{\prime
})}{\partial t^{\prime }}\;,
\end{eqnarray}
with $\Delta A$ of Eq. (20). Because of Eq. (27) we can write Eq.
(42) in the alternative form
%
\begin{eqnarray}
W(t) &=&-\frac{1}{\Delta t}\int_{t}^{t+\Delta t}dt^{\prime }\int
d^{3}r^{\prime }\int d^{3}r^{\prime \prime }\int_{-\infty }^{\infty
}d\tau \,\chi (\mathbf{r}^{\prime },\mathbf{r}^{\prime \prime };\tau
|t^{\prime
})\times   \notag \\
&&\mathcal{F}(\mathbf{r}^{\prime \prime },\tau +t^{\prime
})\frac{\partial \mathcal{F}(\mathbf{r}^{\prime },t^{\prime
})}{\partial t^{\prime }}\;,
\end{eqnarray}
with $\tau =t^{\prime \prime }-t^{\prime }$, and using Eq. (37) and
the Fourier transform of $\chi $, it follows
%
\begin{eqnarray}
W(t) &=&-\frac{1}{\Delta t}\int_{t}^{t+\Delta t}dt^{\prime }\int
d^{3}r^{\prime }\int d^{3}r^{\prime \prime }\,\int d\omega i\omega
\mathcal{F}(\mathbf{r}^{\prime \prime },\omega
)\mathcal{F}(\mathbf{r}^{\prime
},\omega )\times   \notag \\
&&\Bigl\{\chi ^{\ast }(\mathbf{r}^{\prime },\mathbf{r}^{\prime
\prime };\omega |t^{\prime })(e^{2i\omega t^{\prime }}-1)-\chi
(\mathbf{r}^{\prime },\mathbf{r}^{\prime \prime };\omega |t^{\prime
})(e^{-2i\omega t^{\prime }}-1)\Bigr\}\;.
\end{eqnarray}

If we consider a particular situation in which
$\mathcal{R}_{\varepsilon }(t^{\prime })$ varies weakly in time in
the interval $\Delta t$ (implying in that ultrafast relaxation
processes are not present), we can approximate it by
$\mathcal{R}_{\varepsilon }(t)$, and in conditions such that $\Delta
t$ involves several periods $2\pi /\omega $ so that the exponentials
cancel on average, using Eq. (23) it follows that
%
\begin{equation}
W(t)=\frac{1}{2}\int d^{3}r^{\prime }\int d^{3}r^{\prime \prime
}\int d\omega \mathcal{F}(\mathbf{r}^{\prime },\omega
)\mathcal{F}(\mathbf{r}^{\prime \prime },\omega )\alpha
(\mathbf{r}^{\prime },\mathbf{r}^{\prime \prime };\omega |t)\;,
\end{equation}
where
%
\begin{equation}
\alpha (\mathbf{r}^{\prime },\mathbf{r}^{\prime \prime };\omega |t)=\omega
\chi ^{\prime \prime }(\mathbf{r}^{\prime },\mathbf{r}^{\prime \prime
};\omega |t)
\end{equation}
can be considered a kind of absorption coefficient at each frequency $\omega
$ associated to the perturbing force, and at the macroscopic (nonequilibrium
thermodynamic) state of the system at time $t$.

Evidently, and this is a fundamental point to be stressed, the
generalized susceptibility depends on $\varrho_{\varepsilon}$,
through $\mathcal{R}_{\varepsilon}$, and since the latter is a
functional of the time- (and eventually of the space-) dependent
variables that characterized the nonequilibrium thermodynamic state
of the system, then \textit{the calculus of responses needs be
coupled with the one of the equations of evolution for the basic
variables, as given by the nonlinear quantum kinetic theory that the
formalism provides} [8-11,14].

Closing this section let us add some additional considerations.

First, if the system has translational invariance (or near
\textit{translational invariance} as in the case of regular
crystalline matter), the dependence of $\chi ^{\prime \prime }$ and
$\chi $ on $\mathbf{r}$ and $\mathbf{r}^{\prime }$ is through the
difference $\mathbf{r}-\mathbf{r}^{\prime }$. We can then introduce
the Fourier transform in space, namely
%
\begin{equation}
\chi ^{\prime \prime }(\mathbf{k},\omega |t)=\int d^{3}b \,\chi
^{\prime \prime }(b,\omega |t)e^{-i\mathbf{k}\cdot \mathbf{b}}\;,
\end{equation}
where $ \mathbf{b} = \mathbf{r}-\mathbf{r}^{\prime }$. Moreover,
%
\begin{equation}
\lbrack \chi ^{\prime \prime }(\mathbf{r},\mathbf{r}^{\prime };-\tau
|t)]^{\ast }=-\chi ^{\prime \prime }(\mathbf{r},\mathbf{r}^{\prime };-\tau
|t)\;,
\end{equation}
as a result of $\chi ^{\prime \prime }$ involving a commutator of
Hermitian operators, and this $\chi ^{\prime \prime }$ is then a
purely imaginary quantity. Similarly, it follows that
%
\begin{equation}
\chi ^{\prime \prime }(\mathbf{r},\mathbf{r}^{\prime };-\tau |t)=-\chi
^{\prime \prime }(\mathbf{r},\mathbf{r}^{\prime };\tau |t)\;,
\end{equation}
what implies that
%
\begin{equation}
\chi ^{\prime \prime }(\mathbf{k},\omega |t)=-\chi ^{\prime \prime
}(-\mathbf{k},-\omega |t)\;.
\end{equation}

On the other hand, on the basis of Eqs. (32) and (50) we have that
%
\begin{eqnarray}
Im\,\chi (\mathbf{k},\omega |t) &=&\frac{1}{2i}\bigl(\ \chi
(\mathbf{k}
,\omega |t)-\chi ^{\ast }(\mathbf{k},\omega |t)\bigr)  \notag \\
&=&\frac{1}{2i}\int_{-\infty }^{\infty }\frac{d\omega }{\pi }\int
d^{3}b \int d\tau \,\chi ^{\prime \prime }(\mathbf{b},-\tau
|t)\left( \frac{e^{-i( \mathbf{k}\cdot \mathbf{b}-\omega ^{\prime
}\tau )}}{\omega ^{\prime }-\omega -is}+\frac{e^{i(\mathbf{k}\cdot
\mathbf{b}-\omega ^{\prime }\tau )}}{\omega
^{\prime }-\omega +is}\right) \hfill   \notag \\
&=&\frac{1}{2i}\int_{-\infty }^{\infty }\frac{d\omega }{\pi }\int
d^{3}b \int d\tau \,e^{-i(\mathbf{k}\cdot \mathbf{b}-\omega \tau
)}\left( \frac{\chi ^{\prime \prime }(\mathbf{b},-\tau |t)}{\omega
^{\prime }-\omega -is}+\frac{ \chi ^{\prime \prime
}(-\mathbf{b},\tau |t)}{\omega ^{\prime }-\omega +is}
\right) \hfill   \notag \\
&=&\chi ^{\prime \prime }(\mathbf{k},\omega |t)\,,
\end{eqnarray}
which is then a real quantity, and also is $\chi ^{\prime \prime
}(\mathbf{r},\mathbf{r}^{\prime };\omega |t)$.

Moreover, if the initial condition of preparation of the system is
the one of equilibrium with the reservoirs, characterized by
distribution $\varrho _{\text{eq}}$ (which commutes with $\Hat{H}$),
we recover the usual expression [28]
%
\begin{equation}
\chi (\mathbf{r},\mathbf{r}^{\prime };t-t^{\prime
})_{\text{eq}}=\theta (t-t^{\prime })\frac{i}{\hbar
}Tr\bigl\{[\Hat{A}(\mathbf{r}),\tilde{A}(\mathbf{r}^{\prime
},t^{\prime }-t)]\,\varrho _{\text{eq}}\bigr\}\;.
\end{equation}
where, as usual, because the equilibrium has been established, the
interaction of system and reservoir is neglected.

Secondly, as pointed out in the Introduction the notable
developments in instrumentation that are at present being
accumulated, and which are necessary for the study of systems
working in far-from-nonequilibrium conditions in the sought-after
miniaturized devices with ultrafast responses of nowadays advanced
technology, require the mechanical-statistical analysis in short
time intervals (as described above) and in nanometric spatial
regions. This is also contained in the theoretical treatment already
described. In fact, if the property of the systems is measured in a
small region around position $\mathbf{r}$, and also evolving in
time, the expected value is given in Eq. (27), which we can
alternatively write as
\begin{equation}
\Delta \langle \Hat{A}(\mathbf{r})|t\rangle =\frac{1}{2\hbar }\int_{-\infty
}^{\infty }\!\!dt^{\prime }Tr\bigl\{[\hat{A}(\mathbf{r},t),\mathcal{\hat{V}}
(t^{\prime },t^{\prime }-t)]\mathcal{R}_{\varepsilon }(t)\}\bigr\}\;,  \notag
\end{equation}
with [cf. Eq. (2)]
\begin{equation}
\mathcal{\hat{V}}(t^{\prime },t^{\prime }-t)=-\int \!d^{3}r^{\prime }
\mathcal{F}(\mathbf{r}^{\prime },t^{\prime })\hat{A}(\mathbf{r}^{\prime
},t^{\prime }-t)\;,  \notag
\end{equation}

Finally this NESEF-based approach to Response Function Theory
involves, as seen above, the calculation of averages in terms of the
nonequilibrium statistical operator, a quite difficult task. First
we recall that, we have taken the system in contact with external
reservoirs, the latter very much larger than the system, which, for
all practical purposes, remain in a stationary state of equilibrium
all along the realization of the experiments, and that $
\mathcal{R}_{\varepsilon }(t)=\varrho _{\varepsilon }(t)\times
\varrho _{R}$, where $\varrho _{R}$ is the stationary equilibrium
statistical operator of the reservoir(s), and $\varrho _{\varepsilon
}(t)$ the nonequilibrium statistical operator of the system.
Moreover, we can write [8-11]
%
\begin{equation}
\varrho _{\varepsilon }(t)=\bar{\varrho}(t)+\varrho _{\varepsilon }^{\prime
}(t)\;,
\end{equation}
that is, the sum of the "instantaneously-frozen" auxiliary
statistical operator $\bar{\varrho}$ and the contribution $\varrho
_{\varepsilon }^{\prime }$ which accounts for the relaxation
processes developing in the media. Therefore, we can write Eq. (25)
as
%
\begin{equation}
\chi ^{\prime \prime }(\mathbf{r},\mathbf{r}^{\prime };t^{\prime
}-t|t)=\bar{ \chi}^{\prime \prime }(\mathbf{r},\mathbf{r}^{\prime
};t^{\prime }-t|t)+\chi _{\varepsilon }^{\prime \prime
}(\mathbf{r},\mathbf{r}^{\prime };t^{\prime }-t|t)\;.
\end{equation}
where in $\bar{\chi}^{\prime \prime }$ the averaging is over the
auxiliary ensemble, characterized by $\bar{\varrho}$, and $\chi
_{\varepsilon }^{\prime \prime }$ is the averaging in terms of the
contribution $\varrho_{\varepsilon}^{\prime}$. Using the expression
for $\varrho _{\varepsilon}^{\prime}$, given in terms of
$\bar{\varrho}$ [11], it can be shown that it is expressed in a
Born-perturbation-like series in powers of $H^{\prime }$, the
internal interactions in the system. Therefore, whereas the
weak-coupling limit can be used, we can retain only $\bar{\chi}$ to
a good degree of approximation. In the NESEF-based kinetic equations
the use of such limit renders the equation Markovian in character
[14]. We stress again that the \emph{Response Function Theory for
systems away from equilibrium is always coupled with the kinetic
equations that describe the evolution of the nonequilibrium
thermodynamic state of the system}.

Let us next see another important property of the nonequilibrium generalized
susceptibility, namely a \textit{fluctuation-dissipation theorem in
far-from-equilibrium conditions}.


\section{Fluctuation-Dissipation Theorem in Far-From-Equilibrium Conditions}

The fluctuation-dissipation theorem (\textsc{fdt}) -- originally a
relation between the equilibrium fluctuations in a system and the
dissipative response induced by external forces -- provided a major
impetus for the development of discussions of irreversible
processes. A classical particular form seems to have been proved by
Nyquist [34] for the relationship between the thermal noise and the
impedance of a resistor. Derivation from phenomenological points of
view followed, together with stochastic approaches, and finally
entered into the domain of statistical mechanics [35,36]. We here
extend Kubo's approach in order to encompass arbitrary
nonequilibrium conditions.

As noticed those results involve the immediate neighborhood of the
equilibrium, and were very well established. For systems out of
equilibrium (particularly those far from equilibrium) the situation
is not clearly delineated, and some approaches are available for
steady-state conditions in a stochastic approach [37] and in
transient regimes for particular ensembles [38]. We address here the
derivation of a \textsc{fdt} for far-from-equilibrium systems, in
the framework of the nonequilibrium ensemble formalism
\textsc{nesef}, which is a generalization to arbitrary
nonequilibrium conditions of the formalism developed by Kubo [36] in
the case of systems in equilibrium.

A fluctuation-dissipation theorem for systems arbitrarily away from
equilibrium, and in the formalism of \textsc{nesef} here presented, follows
from the comparison of two expressions: one is a correlation function of two
quantities and the other a dynamic response of the system to an external
deterministic perturbation, that is, the generalized susceptibility of the
previous section.

Let us first recall the case of equilibrium [36]. Consider the
quantities $\Hat{A}$ and $\Hat{B}$: their correlation function over
the canonical ensemble in equilibrium is given by
%
\begin{equation}
S_{AB}(\mathbf{r},\mathbf{r}^{\prime },t-t^{\prime })=Tr\{\Delta
\Hat{A}(\mathbf{r},t)\Delta \Hat{B}(\mathbf{r}^{\prime },t^{\prime
})\varrho _{c}\}=Tr\{\Delta \Hat{A}(\mathbf{r},t-t^{\prime
})\Delta \Hat{B}(\mathbf{r} ^{\prime })\varrho _{c}\}\;,
\end{equation}
$\Delta \Hat{A}=\Hat{A}-Tr\{\Hat{A}\varrho _{c}\}$, etc., and the
generalized susceptibility is
%
\begin{equation}
\chi _{AB}^{\prime \prime }(\mathbf{r},\mathbf{r}^{\prime
};t-t^{\prime })= \frac{1}{2\hbar
}Tr\{[\Hat{A}(\mathbf{r},t),\Hat{B}(\mathbf{r}^{\prime },t^{\prime
})]\varrho _{c}\}=\frac{1}{2\hbar }Tr\{[\Hat{A}(\mathbf{r}
,t-t^{\prime }),\Hat{B}(\mathbf{r}^{\prime })]\varrho _{c}\}\;,
\end{equation}
which is a generalization of the one of Eq. (25), with
%
\begin{equation}
\Hat{A}(\mathbf{r},t)=e^{-\frac{1}{i\hbar
}t\Hat{H}}\Hat{A}e^{\frac{1}{i\hbar }t\Hat{H}}\;,
\end{equation}
where $\Hat{H}$ is the Hamiltonian of the system (in the absence of any
external perturbation), that is, the operators are given in Heisenberg
representation, and
%
\begin{equation}
\varrho _{c}=\frac{e^{-\frac{\Hat{H}}{k_{B}T}}}{Z(T,N,V)}
\end{equation}
is the canonical distribution in equilibrium. Using the operational
relationship
%
\begin{equation}
e^{-\beta H}e^{-\frac{1}{i\hbar
}t\Hat{H}}\Hat{A}(\mathbf{r})e^{\frac{1}{ i\hbar
}t\Hat{H}}=e^{-\frac{1}{i\hbar }(t+i\beta \hbar
)H}\Hat{A}(\mathbf{r} )e^{\frac{1}{i\hbar }(t+i\beta \hbar
)\Hat{H}}e^{-\beta \Hat{H}}\equiv \Hat{A}(\mathbf{r},t+i\beta
\hbar )e^{-\beta \Hat{H}}\;,
\end{equation}
we do have that
%
\begin{equation}
Tr\{\varrho _{c}\Hat{A}(\mathbf{r},t)\Hat{B}(\mathbf{r}^{\prime
},t^{\prime })\}=Tr\{\varrho _{c}\Hat{B}(\mathbf{r}^{\prime
},t^{\prime })\Hat{A}(\mathbf{r},t+i\hbar \beta )\}\;.
\end{equation}

Furthermore, let us introduce the Fourier transform in time of the
correlation function, namely
%
\begin{equation}
S_{AB}(\mathbf{r},\mathbf{r}^{\prime };t-t^{\prime
})=\int_{-\infty }^{\infty }\frac{d\omega }{2\pi
}\tilde{S}_{AB}(\mathbf{r},\mathbf{r}^{\prime };\omega )e^{i\omega
(t-t^{\prime })}\;,
\end{equation}
and consider
%
\begin{eqnarray}
\tilde{S}_{AB}(\mathbf{r}^{\prime },\mathbf{r};-\omega )
&=&\int_{-\infty }^{\infty }d\tau \,Tr\{\Delta
\Hat{B}(\mathbf{r}^{\prime })\Delta \Hat{A}(
\mathbf{r},\tau )\varrho _{c}\}e^{-i\omega \tau }  \notag \\
&=&\int_{-\infty }^{\infty }d\tau \,Tr\{\varrho _{c}\Delta
\Hat{B}(\mathbf{r} ^{\prime })\Delta \Hat{A}(\mathbf{r},\tau +i\beta
\hbar )\}\;,
\end{eqnarray}
where we have used the relationship of Eq. (59), and $\tau
=t-t^{\prime }$. Introducing $\tau =\tau ^{\prime }+i\beta \hbar $
it results that
%
\begin{equation}
\tilde{S}_{AB}(\mathbf{r}^{\prime },\mathbf{r},-\omega
)=\tilde{S}_{BA}(\mathbf{r},\mathbf{r}^{\prime };\omega )e^{-\beta
\hbar \omega }\;.
\end{equation}

On the other hand
%
\begin{eqnarray}
\chi _{AB}^{\prime \prime }(\mathbf{r},\mathbf{r}^{\prime };\omega )
&=&\int_{-\infty }^{\infty }d\tau e^{i\omega \tau }\frac{1}{2\hbar
}Tr\{[ \Hat{A}(\mathbf{r},\tau ),\Hat{B}(\mathbf{r}^{\prime
})]\varrho _{c}\}
\notag \\
&=&\int_{-\infty }^{\infty }d\tau e^{i\omega \tau }\frac{1}{2\hbar }
Tr\{[\Delta \Hat{A}(\mathbf{r},\tau ),\Delta
\Hat{B}(\mathbf{r}^{\prime
})]\varrho _{c}\}  \notag \\
&=&\frac{1}{2\hbar }\{\tilde{S}_{AB}(\mathbf{r},\mathbf{r}^{\prime
};\omega )-\tilde{S}_{BA}(\mathbf{r}^{\prime },\mathbf{r};-\omega
)\}\;,
\end{eqnarray}
where we can introduce $\Delta A$ and $\Delta B$ because the extra
terms cancel in the commutation, and using Eq. (63) it follows that
%
\begin{equation}
\tilde{S}_{AB}(\mathbf{r},\mathbf{r}^{\prime };\omega )=2\hbar
\lbrack 1-e^{-\beta \hbar \omega }]^{-1}\chi _{AB}^{\prime \prime
}(\mathbf{r},\mathbf{r}^{\prime };\omega )\;,
\end{equation}
which is the traditional form of the fluctuation-dissipation
theorem. In this condition in the linear regime around equilibrium,
the dependence on the space coordinates is usually of the form
$\mathbf{r}-\mathbf{r}^{\prime }$, and then making the Fourier
transform in the space variable we have that
%
\begin{equation}
\tilde{S}_{AB}(\mathbf{k},\omega )=2\hbar \lbrack 1-e^{-\beta \hbar \omega
}]^{-1}\chi _{AB}^{\prime \prime }(\mathbf{k},\omega )\;,
\end{equation}
where $\chi ^{\prime \prime }$ is the imaginary part of the
generalized susceptibility of the previous Section [cf. Eq. (33)].

Let us now go over the case of a system away from equilibrium, defining
%
\begin{equation}
S_{AB}(\mathbf{r},t;\mathbf{r}^{\prime },t^{\prime }|t_{i})=Tr\{\Delta \Hat{A%
}(\mathbf{r},t)\Delta \Hat{B}(\mathbf{r}^{\prime },t^{\prime
})\varrho _{\varepsilon }(t_{i})\}\;,
\end{equation}
%
\begin{equation}
\chi _{AB}^{\prime \prime }(\mathbf{r},t;\mathbf{r}^{\prime
},t^{\prime
}|t_{i})=\frac{1}{2\hbar }Tr\{[\Hat{A}(\mathbf{r},t),\Hat{B}(\mathbf{r}%
^{\prime },t^{\prime })]\varrho _{\varepsilon }(t_{i})\}\;,
\end{equation}
where $\varrho _{\varepsilon }(t_{i})$ is the distribution
characterizing the preparation, in nonequilibrium conditions, of
the system at time $t_{i}$ when the experiment is initiated, and
$\Delta \Hat{A}(\mathbf{r},t)=\Hat{A}(
\mathbf{r},t)-Tr\{\Hat{A}(\mathbf{r},t)\varrho _{\varepsilon
}(t_{i})\}=\Hat{ A}(\mathbf{r},t)-Tr\{\Hat{A}(\mathbf{r})\varrho
_{\varepsilon }(t)\}$, after using that
%
\begin{equation}
\Hat{A}(\mathbf{r},t)=U^{\dagger }(t)\Hat{A}(\mathbf{r})U(t)\;,
\end{equation}
%
\begin{equation}
\varrho _{\varepsilon }(t)=U(t)\varrho _{\varepsilon
}(t_{i})U^{\dagger }(t)\;,
\end{equation}
%
\begin{equation}
U(t)=e^{-\frac{1}{i\hbar }(t-t_{i})\Hat{H}}\;.
\end{equation}

Further, we write for $\varrho_{\varepsilon}$
%
\begin{equation}
\varrho_{\varepsilon}(t)=\exp\{-\Hat{S}_{\varepsilon}(t)\}\;,
\end{equation}
where
%
\begin{equation}
\Hat{S}_{\varepsilon}(t)=\Hat{S}(t,0)+\hat{\zeta}_{\varepsilon}(t)\;,
\end{equation}
with
%
\begin{equation}
\hat{\zeta}_{\varepsilon}(t)=-\int_{-\infty}^{t}dt^{\prime}\,e^{
\varepsilon(t^{\prime}-t)}
\frac{d}{dt^{\prime}}\Hat{S}(t^{\prime},t^{\prime}-t)\;.
\end{equation}

Proceeding along a similar way as done in the case of equilibrium, we first
take into account that we can write
%
\begin{eqnarray}
Tr\{\Delta \Hat{B}(\mathbf{r},t)\Delta \Hat{A}(\mathbf{r}^{\prime
},t^{\prime })\varrho _{\varepsilon }(t_{i})\} &=&Tr\{U^{\dagger
}(t)\Delta \Hat{B}(\mathbf{r}^{\prime },t)U(t)U^{\dagger }(t)\Delta
\Hat{A}(\mathbf{r}
)U(t)\varrho _{\varepsilon }(t_{i})\}  \notag \\
&=&Tr\{\Delta \Hat{B}(\mathbf{r}^{\prime },t)\Delta
\Hat{A}(\mathbf{r} )\varrho _{\varepsilon }(t)\}\;.
\end{eqnarray}

Moreover
%
\begin{eqnarray}
Tr\{\Hat{A}(\mathbf{r},t)\Hat{B}(\mathbf{r}^{\prime },t^{\prime
})\varrho _{\varepsilon }(t_{i})\} &=&Tr\{\Hat{B}(\mathbf{r}^{\prime
},t^{\prime
})\varrho _{\varepsilon }(t_{i})\Hat{A}(\mathbf{r},t)\}  \notag \\
&=&Tr\bigl\{e^{-\Hat{S}_{\varepsilon
}(t_{i})}e^{\Hat{S}_{\varepsilon }(t_{i})}U^{\dagger }(t^{\prime
})\Hat{B}(\mathbf{r}^{\prime })U(t^{\prime
})e^{-\Hat{S}_{\varepsilon }(t_{i})}U^{\dagger
}(t)\Hat{A}(\mathbf{r})U(t)
\bigr\}  \notag \\
&=&Tr\bigl\{e^{-\Hat{S}_{\varepsilon }(t_{i})}U^{\dagger
}(t)U(t)e^{\Hat{S} _{\varepsilon }(t_{i})}U^{\dagger
}(t)U(t)U^{\dagger }(t^{\prime })\times
\notag \\
&&\Hat{B}(\mathbf{r}^{\prime })U^{\dagger }(t^{\prime })U^{\dagger
}(t)U(t)e^{-\Hat{S}_{\varepsilon }(t_{i})}U^{\dagger
}(t)\Hat{A}(\mathbf{r}
)U(t)\}  \notag \\
&=&Tr\biggl\{\bigl(U(t)e^{-\Hat{S}_{\varepsilon }(t_{i})}U^{\dagger
}(t) \bigr)\bigl(U(t)e^{\Hat{S}_{\varepsilon }(t_{i})}U^{\dagger
}(t)\bigr)\times
\notag \\
&&\bigl(U(t)U^{\dagger }(t^{\prime })\Hat{B}(\mathbf{r}^{\prime
})U(t^{\prime })U^{\dagger }\bigr)\bigl(U(t)e^{-\Hat{S}_{\varepsilon
}(t_{i})}U^{\dagger }(t)\bigr)\Hat{A}(\mathbf{r})\biggr\}  \notag \\
&=&Tr\{e^{-\Hat{S}_{\varepsilon }(t)}e^{\Hat{S}_{\varepsilon
}(t)}\Hat{B}( \mathbf{r}^{\prime },t^{\prime
}-t)e^{-\Hat{S}_{\varepsilon }(t)}\Hat{A}(
\mathbf{r})\}  \notag \\
&=&Tr\{\Hat{B}_{\varepsilon }(\mathbf{r}^{\prime },t^{\prime
}-t|t)\Hat{A}( \mathbf{r})\varrho _{\varepsilon }(t)\}\;,
\end{eqnarray}
where
%
\begin{equation}
\Hat{B}_{\varepsilon }(\mathbf{r}^{\prime },-\tau |t)=e^{\Hat{S}
_{\varepsilon }(t)}U(\tau )\Hat{B}(\mathbf{r}^{\prime })U^{\dagger
}(\tau )e^{-\Hat{S}_{\varepsilon }(t)}\;,
\end{equation}
with $\tau =t-t^{\prime }$, and we recall that $U(-\tau )=U^{\dagger }(\tau )
$.

Consider now the susceptibility, when we have that
%
\begin{eqnarray}
\chi _{AB}^{\prime \prime }(\mathbf{r},t;\mathbf{r}^{\prime
},t^{\prime }|t_{i}) &=&\frac{1}{2\hbar
}Tr\{[\Hat{A}(\mathbf{r},t),\Hat{B}(\mathbf{r}
^{\prime },t^{\prime })]\varrho _{\varepsilon }(t_{i})\}  \notag \\
&=&\frac{1}{2\hbar }Tr\{[\Delta \Hat{A}(\mathbf{r},t),\Delta
\Hat{B}(\mathbf{
\ r}^{\prime },t^{\prime })]\varrho _{\varepsilon }(t_{i})\}  \notag \\
&=&\frac{1}{2\hbar }Tr\Bigl\{\bigl(\Delta
\Hat{A}(\mathbf{r},t)\Delta \Hat{B} (\mathbf{r}^{\prime },t^{\prime
})-\Delta \Hat{B}(\mathbf{r}^{\prime },t^{\prime })\Delta
\Hat{A}(\mathbf{r},t)\bigr)\varrho _{\varepsilon
}(t_{i})\Bigr\}  \notag \\
&=&\frac{1}{2\hbar }Tr\{\bigl(\Delta \Hat{B}_{\varepsilon
}(\mathbf{r} ^{\prime },-\tau |t)\Delta \Hat{A}(\mathbf{r})-\Delta
\Hat{B}(\mathbf{r} ^{\prime },-\tau )\Delta
\Hat{A}(\mathbf{r})\bigr)\varrho _{\varepsilon }(t)\}\,\,,
\end{eqnarray}
where we have used Eq. (76), and it can be noticed that the
distribution $\varrho _{\varepsilon }$ in the last expression can be
given as given at time $t$ when a measurement is performed. Equation
(78) can now be written in the form
%
\begin{equation}
\chi _{AB}^{\prime \prime }(\mathbf{r},\mathbf{r}^{\prime
};t^{\prime }-t|t)= \frac{1}{2\hbar }[S_{BA}^{\varepsilon
}(\mathbf{r}^{\prime },\mathbf{r};t^{\prime
}-t|t)-S_{BA}(\mathbf{r}^{\prime },\mathbf{r};t^{\prime }-t|t)]\;,
\end{equation}
after defining
%
\begin{equation}
S_{BA}^{\varepsilon }(\mathbf{r}^{\prime },\mathbf{r};t^{\prime
}-t|t)=Tr\{\Delta \Hat{B}_{\varepsilon }(\mathbf{r}^{\prime },-\tau
)\Delta \Hat{A}(\mathbf{r})\varrho _{\varepsilon }(t)\}\;,
\end{equation}
%
\begin{equation}
S_{BA}(\mathbf{r}^{\prime },\mathbf{r};t^{\prime }-t|t)=Tr\{\Delta \Hat{B}(%
\mathbf{r}^{\prime },-\tau )\Delta \Hat{A}(\mathbf{r})\varrho
_{\varepsilon }(t)\}\;.
\end{equation}

Taking into account that the nonequilibrium distribution admits a
separation consisting of the addition of two parts, one is the
so-called relevant part, $\bar{\varrho}$, plus a contribution,
$\varrho _{\varepsilon }^{\prime }$, which accounts for relaxation
processes which are governed by $\Hat{H}^{\prime }$ in Eq. (1), we
separate the fluctuation-dissipation relation of Eq. (64) in a
\textquotedblleft relevant\textquotedblright\ part (the one
depending on $\bar{\varrho}$ and $H_{0}$ alone) and the rest. That
\textquotedblleft relevant\textquotedblright\ part is then
%
\begin{eqnarray}
\bar{\chi}_{AB}^{\prime \prime }(\mathbf{r},\mathbf{r}^{\prime
},-\tau |t) &=&\frac{1}{2\hbar
}Tr\Bigl\{\bigl(e^{\Hat{S}(t,0)}\Delta \Hat{B}(\mathbf{r}
^{\prime },-\tau )_{0}e^{-\Hat{S}(t,0)}\Delta \Hat{A}(\mathbf{r})-  \notag \\
&&\Delta \Hat{B}(\mathbf{r}^{\prime },\tau )_{0}\Delta
\Hat{A}(\mathbf{r}) \bigr)\bar{\varrho}(t,0)\Bigr\}\;,
\end{eqnarray}
where
%
\begin{equation}
\Delta \Hat{B}(\mathbf{r},-\tau )_{0}=e^{\frac{1}{i\hbar }\tau
\Hat{H}_{0}}\Hat{B}(\mathbf{r})e^{-\frac{1}{i\hbar }\tau
\Hat{H}_{0}}\;.
\end{equation}

Transforming Fourier in $\tau $ and also in the space coordinates
after assuming dependence on $\mathbf{r}-\mathbf{r}^{\prime }$, that
is, in the cases when the system displays translational invariance,
we have that Eq. (79) becomes
%
\begin{equation}
\chi _{AB}^{\prime \prime }(\mathbf{k},\omega |t)=\frac{1}{2\hbar
} [S_{BA}^{\varepsilon }(\mathbf{k},\omega
|t)-S_{BA}(\mathbf{k},\omega |t)]\;,
\end{equation}
constituting a fluctuation-dissipation relation for systems
arbitrarily deviated from equilibrium. In particular, it goes over
to the one in equilibrium [cf. Eq. (66)] when $\varrho _{\varepsilon
}(t)$ is substituted by the canonical distribution $\varrho _{c}$,
as it should.

On the other hand, in the case of space- and time resolved
experiments in systems with no translational invariance then for
the response in region $\Delta \mathbf{r}$ (the experimental space
resolution with, say, micrometer or nanometric scale) around
position $\mathbf{r}$, we do have
%
\begin{eqnarray}
\chi _{AB\ell }^{\prime \prime }(\mathbf{r};t^{\prime }-t|t) &=&\int
\!d^{3}r^{\prime }\chi _{AB}^{\prime \prime }(\mathbf{r},\mathbf{r}^{\prime
};t^{\prime }-t|t)=  \notag \\
&=&\frac{1}{2\hbar }[S_{BA\ell }^{\varepsilon }(\mathbf{r};t^{\prime
}-t|t)-S_{BA\ell }(\mathbf{r};t^{\prime }-t|t)]\;,
\end{eqnarray}
(with $\ell $ for local), with the last two correlations over the
nonequilibrium ensemble corresponding to the integration over
$\mathbf{r}^{\prime }$ of those on the right of Eq. (79).

We illustrate these results for a specific case, say, a gas of free fermions
for which
%
\begin{equation}
\Hat{H}_{0}=\sum_{\mathbf{k}}\epsilon _{\mathbf{k}}c_{\mathbf{k}}^{\dagger
}c_{\mathbf{k}}\;,
\end{equation}
where $c$($c^{\dagger }$) are the annihilation (creation) operators
in state $\mathbf{k}$ (we omit the spin index), and $\epsilon
_{\mathbf{k}}$ is the energy-dispersion relation. We consider the
case of the nonequilibrium generalized grand-canonical ensemble
[11], when the set of basic variables present in the nonequilibrium
statistical operator are the densities of energy and of particles,
the flux of both (currents), and all the other higher-order
tensorial fluxes of order $r=2,3,\dots $, with $r$ being also the
tensor rank. Further, we restrict the analysis to the homogeneous
condition, i.e. densities and fluxes do not depend on the space
coordinate, and then, in reciprocal space,
%
\begin{equation}
\Hat{S}(t,0)=\sum_{\mathbf{k}}\Biggl\{F_{h}(t)\epsilon _{\mathbf{k}%
}+F_{n}(t)+\sum_{r\geq 1}[F_{h}^{[r]}(t)\otimes
u^{[r]}(\mathbf{k})\epsilon _{\mathbf{k}}+F_{n}^{[r]}(t)\otimes
u^{[r]}(\mathbf{k})]\Biggr\}c_{\mathbf{k}}^{\dagger
}c_{\mathbf{k}}
\end{equation}
is the informational entropy operator in this case.

In this Eq. (87) the $F$'s are the intensive nonequilibrium
thermodynamic variables, associated to the energy, the number of
particles, and the fluxes of these two of order $r$ ($=1,2,\dots $);
$u^{[r]}(\mathbf{k})$ is the tensor consisting of the tensorial
product of $r$ times the generating velocity
$\mathbf{u}(\mathbf{k})=\hbar ^{-1}\nabla _{\mathbf{k}}\epsilon _{
\mathbf{k}}$, namely the group velocity of the fermion in state
$\mathbf{k}$; sign $\otimes $ stands for fully contracted product of
tensors.

Equation (87) can be written in the compact form
%
\begin{equation}
\Hat{S}(t,0)=\sum_{\mathbf{k}}\varphi
_{\mathbf{k}}(t)c_{\mathbf{k}}^{\dagger }c_{\mathbf{k}}\;,
\label{TFReq90}
\end{equation}
where, evidently, $\varphi$ is the quantity between curly brackets
in Eq. (87). Hence, in these conditions we have that the
\textquotedblleft relevant\textquotedblright\ part of the
correlation of Eq. (80) is
%
\begin{eqnarray}
S_{BA}^{\varepsilon }(\mathbf{r}^{\prime },\mathbf{r};-\tau
|t)^{\text{rel}} &=&Tr\Biggl\{e^{\Hat{S}(t,0)}e^{-\frac{1}{i\hbar
}\tau \Hat{H}_{0}}\Hat{B}( \mathbf{r}^{\prime })e^{\frac{1}{i\hbar
}\tau \Hat{H}_{0}}e^{-\Hat{S}(t,0)}
\Hat{A}(\mathbf{r})\bar{\varrho}(t,0)\Biggr\}  \notag \\
&=&Tr\Biggl\{e^{\frac{1}{i\hbar }\sum_{\mathbf{k}}\tau
_{\mathbf{k}}\epsilon _{\mathbf{k}}c_{\mathbf{k}}^{\dagger
}c_{\mathbf{k}}}\Hat{B}(\mathbf{r} ^{\prime })e^{-\frac{1}{i\hbar
}\sum_{\mathbf{k}}\tau _{\mathbf{k}}\epsilon
_{\mathbf{k}}c_{\mathbf{k}}^{\dagger
}c_{\mathbf{k}}}\Hat{A}(\mathbf{r})\bar{ \varrho}(t,0)\Biggr\}\;,
\end{eqnarray}
where we have defined the quantity
%
\begin{equation}
\tau _{\mathbf{k}}=\tau -i\hbar (\varphi _{\mathbf{k}}(t)/\epsilon
_{\mathbf{k}})\;,
\end{equation}
which has dimension of time.

Let us further simplify matters and use a truncation in the basic
set of macrovariables, retaining only the energy $\Hat{H}_{0}$, the
number of particles $N$, and the flux of matter $\mathbf{I}_{n}$,
which multiplied by the mass of the fermions becomes the linear
momentum, $\mathbf{P}$. We call the associated nonequilibrium
thermodynamic variables $F_{h}=\beta ^{\ast }(t)$, $F_{n}=-\beta
^{\ast }(t)\mu ^{\ast }(t)$, and $\mathbf{F}_{n}(t)=-\beta ^{\ast
}(t)m^{\ast }\mathbf{v}(t)$, introducing $\beta ^{\ast
-1}(t)=k_{B}T^{\ast }(t)$, a reciprocal of a quasitemperature, $\mu
^{\ast }(t)$ a quasi-chemical potential, and a drift velocity
$\mathbf{v}(t)$ [11,39-42]. This implies in that we are using a kind
of nonequilibrium canonical distribution with an additional term
arising out of the presence of the current with drift velocity
$\mathbf{v}(t)$. Moreover, let us choose for $\Hat{A}$ and $\Hat{B}$
the nondiagonal elements of Dirac-Landau-Wigner single-particle
density matrix written in second quantization, namely
%
\begin{equation}
\Hat{A}\equiv \hat{n}_{\mathbf{kQ}}^{\dagger
}=c_{\mathbf{k}-\frac{1}{2} \mathbf{Q}}^{\dagger
}c_{\mathbf{k}+\frac{1}{2}\mathbf{Q}}\;,
\end{equation}
%
\begin{equation}
\Hat{B}\equiv
\hat{n}_{\mathbf{kQ}}=c_{\mathbf{k}+\frac{1}{2}\mathbf{Q} }^{\dagger
}c_{\mathbf{k}-\frac{1}{2}\mathbf{Q}}\;,
\end{equation}
which will appear in the calculation of inelastic scattering cross
sections later on. Hence the relevant part of the corresponding
correlation is
%
\begin{equation}
S_{nn^{\dagger }}(\mathbf{k},\mathbf{Q},-\tau |t)^{\text{rel}
}=Tr\{\tilde{U}(\tau |t)\hat{n}_{\mathbf{kQ}}\tilde{U}^{\dagger
}(\tau |t) \hat{n}_{\mathbf{kQ}}^{\dagger }\bar{\varrho}(t,0)\}\;,
\end{equation}
where
%
\begin{equation}
\tilde{U}(\tau |t)=\exp \Biggl\{\frac{1}{i\hbar
}\sum_{\mathbf{k}}\Bigl[ \epsilon _{\mathbf{k}}(\tau -i\hbar \beta
^{\ast }(t))-i\hbar \beta ^{\ast }(t)\mu ^{\ast }(t)+i\hbar \beta
^{\ast }(t)\mathbf{v}(t)\cdot \hbar \mathbf{
k}\Bigr]c_{\mathbf{k}}^{\dagger }c_{\mathbf{k}}\Biggr\}\;,
\end{equation}
and we have that
%
\begin{eqnarray}
\tilde{U}(\tau |t)n_{\mathbf{kQ}}\tilde{U}^{\dagger }(\tau |t)
&=&\exp \biggl\{\frac{1}{i\hbar }(-\tau +i\hbar \beta ^{\ast
}(t))(\epsilon _{ \mathbf{k}+\frac{1}{2}\mathbf{Q}}-\epsilon
_{\mathbf{k}-\frac{1}{2}\mathbf{Q} })+i\hbar \beta ^{\ast
}(t)\mathbf{v}(t)\cdot \mathbf{Q}\biggr\}n_{\mathbf{kQ
}}  \notag \\
&=&e^{\beta ^{\ast }(t)\mathbf{v}(t)\cdot
\mathbf{Q}}n_{\mathbf{kQ}}(-\tau +i\hbar \beta ^{\ast }(t))
\end{eqnarray}

Hence
%
\begin{equation}
S_{nn^{\dagger }}(\mathbf{k},\mathbf{Q};-\tau
|t)^{\text{rel}}=e^{\hbar \beta (t)\mathbf{v}(t)\cdot
\mathbf{Q}}Tr\{n_{\mathbf{kQ}}(-\tau +i\hbar \beta ^{\ast
}(t))n_{\mathbf{kQ}}^{\dagger }\bar{\varrho}(t,0)\}\;,
\end{equation}
and using Eq. (82), after some calculation we find that
%
\begin{equation}
S_{nn^{\dagger }}(\mathbf{k},\mathbf{Q};-\tau |t)^{\text{rel}
}=2\hbar \lbrack 1-e^{-\beta (\hbar \omega -\mathbf{v}(t)\cdot
\mathbf{Q} )}]^{-1}\chi _{nn^{\dagger
}}(\mathbf{k},\mathbf{Q};\omega |t)^{\text{rel}}
\end{equation}

For $\mathbf{v}=0$ and in the case of equilibrium we recover the
well known result of Eq. (66).


\section{Nonequilibrium-Thermodynamic Green Functions}

According to the previous sections, to obtain response functions
requires the calculation of nonequilibrium correlation functions.
This is a difficult mathematical task which can be facilitated by
the introduction of appropriate nonequilibrium thermodynamic Green
functions [43-45]. The approach is an extension of the equilibrium
thermodynamic Green function formalism of Tyablikov and Bogoliubov
[46].

We define the retarded and advanced nonequilibrium-thermodynamic Green
functions of two operators $\Hat{A}$ and $\Hat{B}$ given in Heisenberg
representation, by the expressions
%
\begin{subequations}
\begin{equation}
\langle \langle \Hat{A}(\mathbf{r},\tau );\Hat{B}(\mathbf{r}^{\prime
})|t\rangle \rangle _{\eta }^{(r)}=\frac{1}{i\hbar }\theta (\tau
)Tr\bigl\{ \lbrack \Hat{A}(\mathbf{r},\tau
),\Hat{B}(\mathbf{r}^{\prime })]_{\eta } \mathcal{R}_{\varepsilon
}(t)\bigr\}\;,
\end{equation}
%
\begin{equation}
\langle \langle \Hat{A}(\mathbf{r},\tau );\Hat{B}(\mathbf{r}^{\prime
})|t\rangle \rangle _{\eta }^{(a)}=-\frac{1}{i\hbar }\theta (-\tau
)Tr\bigl\{ \lbrack \Hat{A}(\mathbf{r},\tau
),\Hat{B}(\mathbf{r}^{\prime })]_{\eta } \mathcal{R}_{\varepsilon
}(t)\bigr\}\;,
\end{equation}
\end{subequations}
where $\tau =t^{\prime }-t$ and $\eta =+$ or $\eta =-$ stands for
anticommutator or commutator of operators $\Hat{A}$ and $\Hat{B}$.
These Green functions satisfy the equations of motion
%
\begin{equation}
i\hbar \frac{\partial }{\partial \tau }\langle \langle \Hat{A}(\mathbf{r}
,\tau );\Hat{B}(\mathbf{r}^{\prime })|t\rangle \rangle _{\eta
}^{(r,a)}=\delta (\tau )Tr\bigl\{[\Hat{A}(\mathbf{r}),\Hat{B}(\mathbf{r}
^{\prime })]_{\eta }\mathcal{R}_{\varepsilon }(t)\bigr\}+\langle \langle
\lbrack \Hat{A}(\mathbf{r},\tau ),H];\Hat{B}(\mathbf{r}^{\prime })|t\rangle
\rangle _{\eta }^{r,a}\;.
\end{equation}

In Eq. (99), and in what follows,
$[\Hat{A}(\mathbf{r}),\Hat{B}(\mathbf{r}^{\prime })]$ without
subscript is the commutator of quantities $A$ and $B$. Introducing
the Fourier transform
%
\begin{equation}
\langle \langle \Hat{A}(\mathbf{r});\Hat{B}(\mathbf{r}^{\prime })|\omega
;t\rangle \rangle _{\eta }=\int_{-\infty }^{\infty }d\tau \exp [i\omega \tau
]\langle \langle \Hat{A}(\mathbf{r},\tau );\Hat{B}(\mathbf{r}^{\prime
})|t\rangle \rangle _{\eta }\; ,
\end{equation}
equation (99) becomes
%
\begin{equation}
\hbar \omega \langle \langle
\Hat{A}(\mathbf{r});\Hat{B}(\mathbf{r}^{\prime })|\omega ;t\rangle
\rangle _{\eta }=\frac{1}{2\pi }Tr\bigl\{[\Hat{A}(
\mathbf{r}),\Hat{B}(\mathbf{r}^{\prime })]_{\eta
}\mathcal{R}_{\varepsilon }(t)\bigr\}+\langle \langle \lbrack
\Hat{A}(\mathbf{r}),H];\Hat{B}(\mathbf{r}^{\prime })|\omega
;t\rangle \rangle _{\eta }\;
\end{equation}
%


\subsection{Green Functions and the Fluctuation-Dissipation Theorem}

Next, we establish the connection of these Green functions with correlation
functions. Consider the nonequilibrium correlation functions
%
\begin{subequations}
\begin{equation}
F_{AB}(\mathbf{r},\mathbf{r}^{\prime };\tau
;t)=Tr\bigl\{\Hat{A}(\mathbf{r} ,\tau )\Hat{B}(\mathbf{r}^{\prime
})\mathcal{R}_{\varepsilon }(t)\bigr\}\;,
\end{equation}
%
\begin{equation}
F_{BA}(\mathbf{r},\mathbf{r}^{\prime };\tau
;t)=Tr\bigl\{\Hat{B}(\mathbf{r} ^{\prime })\Hat{A}(\mathbf{r},\tau
)\mathcal{R}_{\varepsilon }(t)\bigr\}\;,
\end{equation}
\end{subequations}
and let $|n\rangle $ and $E_{n}$ be the eigenstates and
eigenvalues of the Hamiltonian $\Hat{H}$. Defining the
nonequilibrium spectral density functions
%
\begin{subequations}
\begin{equation}
J_{AB}(\mathbf{r},\mathbf{r}^{\prime };\omega |t)=2\pi
\sum_{lmn}\langle n| \Hat{A}(\mathbf{r}^{\prime })|m\rangle \langle
m|\Hat{B}(\mathbf{r}^{\prime })|l\rangle \langle
l|\mathcal{R}_{\varepsilon }(t)|n\rangle \delta (\hbar \omega
-E_{m}+E_{n})\;,
\end{equation}
%
\begin{equation}
K_{BA}(\mathbf{r},\mathbf{r}^{\prime };\omega |t)=2\pi
\sum_{lmn}\langle n| \Hat{B}(\mathbf{r}^{\prime })|m\rangle \langle
m|\Hat{A}(\mathbf{r}^{\prime })|l\rangle \langle
l|\mathcal{R}_{\varepsilon }(t)|n\rangle \delta (\hbar \omega
-E_{l}+E_{m})\;,
\end{equation}
\end{subequations}
we obtain the relations
%
\begin{subequations}
\begin{equation}
F_{AB}(\mathbf{r},\mathbf{r}^{\prime };\tau ;t)=\int_{-\infty
}^{\infty } \frac{d\omega }{2\pi
}J_{AB}(\mathbf{r},\mathbf{r}^{\prime };\omega |t)\exp [-i\omega
\tau ]\;,
\end{equation}
%
\begin{equation}
\tilde{F}_{BA}(\mathbf{r},\mathbf{r}^{\prime };\tau
;t)=\int_{-\infty }^{\infty }\frac{d\omega }{2\pi
}K_{BA}(\mathbf{r},\mathbf{r}^{\prime };\omega |t)\exp [-i\omega
\tau ]\;,
\end{equation}
\end{subequations}
and
%
\begin{subequations}
\begin{equation}
\langle \langle \Hat{A}(\mathbf{r}^{\prime
});\Hat{B}(\mathbf{r}^{\prime })|\omega \pm is;t\rangle \rangle
_{+}+\langle \langle \Hat{A}(\mathbf{r} ^{\prime
});\Hat{B}(\mathbf{r}^{\prime })|\omega \pm is;t\rangle \rangle
_{-}=\frac{1}{\hbar }\int_{-\infty }^{\infty }\frac{d\omega ^{\prime
}}{\pi } \frac{J_{AB}(\mathbf{r},\mathbf{r}^{\prime };\omega
^{\prime }|t)}{\omega -\omega ^{\prime }\pm is}\;,
\end{equation}
%
\begin{equation}
\langle \langle \Hat{A}(\mathbf{r}^{\prime
});\Hat{B}(\mathbf{r}^{\prime })|\omega \pm is;t\rangle \rangle
_{-}-\langle \langle \Hat{A}(\mathbf{r} ^{\prime
});\Hat{B}(\mathbf{r}^{\prime })|\omega \pm is;t\rangle \rangle
_{+}=\frac{1}{\hbar }\int_{-\infty }^{\infty }\frac{d\omega ^{\prime
}}{\pi } \frac{K_{BA}(\mathbf{r},\mathbf{r}^{\prime };\omega
^{\prime }|t)}{\omega -\omega ^{\prime }\pm is}\;,
\end{equation}
\end{subequations}
with $s\rightarrow +\!0$, plus sign is for the retarded and minus
sign for the advanced Green functions, and we made use of the
relation
%
\begin{equation}
\int_{-\infty }^{\infty }d\tau \,\theta (\pm \tau )\exp [i\,(\omega -\omega
^{\prime })\tau ]=\pm \frac{i}{\omega -\omega ^{\prime }\pm is}\;.
\end{equation}

Equations (105) may be considered particular generalizations of the
fluctuation-dissipation theorem for systems arbitrarily away from
equilibrium. Near equilibrium, replacing $\varrho _{\varepsilon }$
by the canonical Gibbs distribution we recover the well known result
%
\begin{equation}
\langle \langle \Hat{A}(\mathbf{r}^{\prime });\Hat{B}(\mathbf{r}^{\prime
})|\omega +is\rangle \rangle _{\eta }^{\text{equil.}}-\langle \langle \Hat{A}
(\mathbf{r}^{\prime });\Hat{B}(\mathbf{r}^{\prime })|\omega -is\rangle
\rangle _{\eta }^{\text{equil.}}=\frac{1}{i\hbar }(1-\eta \exp [-\beta \hbar
\omega ])J_{AB}^{\text{equil.}}(\omega )\;,
\end{equation}
where $\beta =(k_{B}T)^{-1}$.

We recall that the nonequilibrium-thermodynamic Green functions of
Eqs. (98) depend on the macroscopic state of the system, and
therefore their equations of motion, Eqs. (99) or (101), must be
solved coupled to the generalized nonlinear transport equations for
the basic set of nonequilibrium thermodynamic variables [14].
Finally, if we write for the interaction energy $\mathcal{V}=\lambda
\exp [i\omega t]\Hat{B}(\mathbf{r}^{\prime })$ where $\lambda $ is a
coupling strength constant, it follows that
%
\begin{eqnarray}
\langle \Hat{A}(\mathbf{r})|t\rangle -\langle
\Hat{A}(\mathbf{r})|t\rangle ^{0} &=&-\frac{\lambda }{i\hbar
}\int_{-\infty }^{0}d\tau \exp [-i\omega \tau
]Tr\{[\Hat{A}(\mathbf{r}),\tilde{B}(\mathbf{r}^{\prime },\tau )]
\mathcal{R}_{\varepsilon }(t)\}+\text{c.c.}  \notag \\
&=&-\frac{\lambda }{i\hbar }\int_{-\infty }^{\infty }d\tau \theta
(\tau )[
\tilde{F}_{\tilde{A}\tilde{B}}(\mathbf{r},\mathbf{r}^{\prime };\tau
;t)-F_{ \tilde{B}\tilde{A}}(\mathbf{r},\mathbf{r}^{\prime };\tau
;t)]\exp [i\omega
\tau ]+\text{c.c.}  \notag \\
&=&\frac{\lambda }{\hbar }\int_{-\infty }^{\infty }\frac{d\omega
^{\prime }}{ 2\pi
}\frac{K_{\tilde{A}\tilde{B}}(\mathbf{r},\mathbf{r}^{\prime };\omega
^{\prime }|t)-J_{\tilde{B}\tilde{A}}(\mathbf{r},\mathbf{r}^{\prime
};\omega
^{\prime }|t)}{\omega -\omega ^{\prime }+is}+\text{c.c.}  \notag \\
&=&2\lambda Re\langle \langle \tilde{B}(\mathbf{r}^{\prime
});\tilde{A}( \mathbf{r})|\omega +is;t\rangle \rangle _{-}\;,
\end{eqnarray}
where $Re$ stands for real part, and we have used the definition of
the advanced Green function of Eq. (98). Hence, the linear response
function to an external harmonic perturbation is given by an
advanced nonequilibrium-thermodynamic Green function dependent on
the macroscopic state of the system characterized by the
nonequilibrium thermodynamic macrovariables $F_{j}(t)$ [or
equivalently $Q_{j}(t)$], as described in Section II.

Closing this section we note that since the nonequilibrium
thermodynamic Green functions of Eqs. (98) are defined as
nonequilibrium averages of dynamical quantities, recalling the
separation of $\varrho _{\varepsilon }$ in a secular and non-secular
(dissipative) parts, we can write
%
\begin{equation}
\langle \langle \Hat{A}(\mathbf{r}^{\prime
});\Hat{B}(\mathbf{r}^{\prime })|\omega ;t\rangle \rangle =\langle
\langle \Hat{A}(\mathbf{r}^{\prime}); \Hat{B}(\mathbf{r}^{\prime
})|\omega ;t\rangle \rangle ^{\text{sec.} }+\langle \langle
\Hat{A}(\mathbf{r}^{\prime });\Hat{B}(\mathbf{r}^{\prime })|\omega
;t\rangle \rangle ^{\prime }\;,
\end{equation}
where
%
\begin{subequations}
\begin{equation}
\langle \langle \Hat{A}(\mathbf{r}^{\prime
});\Hat{B}(\mathbf{r}^{\prime })|\omega ;t\rangle \rangle
^{\text{sec.}}=\pm \frac{1}{i\hbar } \int_{-\infty }^{\infty }d\tau
\exp [i\omega \tau ]\theta (\pm \tau )Tr\{[
\Hat{A}(\mathbf{r}^{\prime },\tau ),\Hat{B}(\mathbf{r}^{\prime
})]_{\eta } \bar{\mathcal{R}}(t,0)\}\;,
\end{equation}
%
\begin{equation}
\langle \langle \Hat{A}(\mathbf{r}^{\prime
});\Hat{B}(\mathbf{r}^{\prime })|\omega ;t\rangle \rangle ^{\prime
}=\pm \frac{1}{i\hbar }\int_{-\infty }^{\infty }d\tau \exp [i\omega
\tau ]\theta (\pm \tau )\{[\Hat{A}(\mathbf{r} ^{\prime },\tau
),\Hat{B}(\mathbf{r}^{\prime })]_{\eta };\mathcal{R} _{\varepsilon
}^{\prime }(t)|t\}\;.
\end{equation}
\end{subequations}

We notice that, in general, the last term in Eq. (101) couples the
equation for the Green function with higher order Green functions,
whose equations must be written and thus one obtains a hierarchy of
coupled equations. Usually one solves this hierarchy of equations
introducing a truncation procedure, like some kind of random phase
approximation. For these nonequilibrium-thermodynamic Green
functions \textit{a second type of expansion and truncation} is also
present, which is that associated with the irreversible processes
encompassed in the contribution
$\mathcal{R}^{\prime}_{\varepsilon}(t)$ to the statistical operator
present in Eq. (110b). Care should be taken to perform consistently
both types of truncation procedures, i.e. to maintain terms of the
same order in the interaction strengths. We recall that the
Markovian approximation in the \textsc{nesef}-based kinetic theory
[14,47] requires to keep terms containing the operator for the
interaction energies up to second order only. The formalism of this
section was applied to the study of time-resolved Raman spectroscopy
as described in Refs. [48,49].


\section{Theory of Scattering for Far-From-Equilibrium Systems}

In a scattering experiment a beam of particles (e.g. photons, ions,
electrons, neutrons, etc.) with, say, energy $\varepsilon _{0}$ and
momentum $\hbar \mathbf{k}_{0}$, incide on a sample, where they
interact with one or more subsystems of it (say, atoms, molecules,
electrons, phonons, etc.). The particles are scattered, as a result
of that interaction, involving a transference of energy $\Delta E$,
and of momentum $\hbar \mathbf{q}$, consequence of an excitation
being created or annihilated in the system; Fig. 2 shows a scheme of
the experiment.

%
\begin{figure}[h]
\center
\includegraphics[width=10cm]{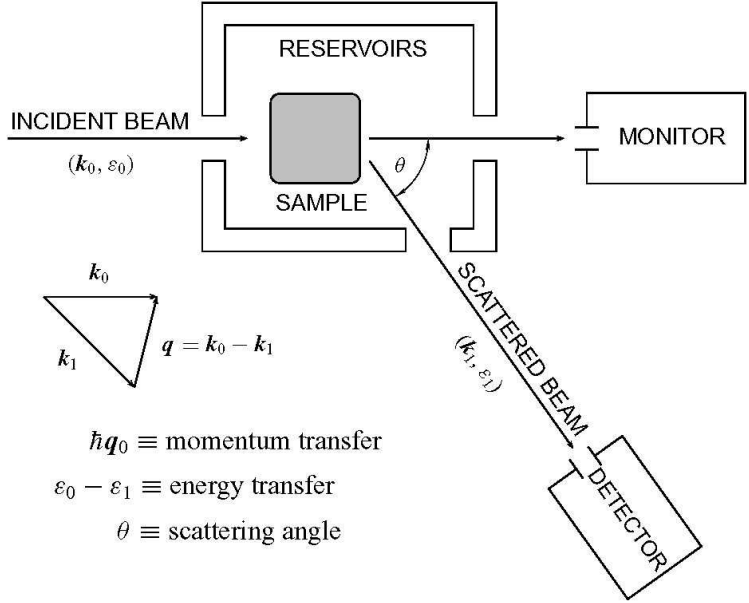}
\caption{Scheme of an experiment of scattering.}
\end{figure}

Calling $\epsilon _{1}$ and $\hbar \mathbf{k}_{1}$ the energy and momentum
of the scattered particle, conservation of energy and momentum require that
%
\begin{equation}
\Delta E =\varepsilon _{0}-\varepsilon _{1}\;,
\end{equation}
%
\begin{equation}
\hbar \mathbf{q} =\hbar \mathbf{k}_{0}-\hbar \mathbf{k}_{1}\;,
\end{equation}
or
%
\begin{equation}
q^{2}=k_{0}^{2}+k_{1}^{2}-2k_{0}k_{1}\cos \theta
\end{equation}
after scalar product of Eq. (112) with itself ($\theta $ is the
so-called scattering angle, see Fig. 2).

Let us go over the general theory: The scattering can be characterized by
the quantity \textit{differential scattering cross section,} $d^{2}\sigma
(\Delta E,\mathbf{q})$. It is defined as the ratio between the number of
scattered particles that are collected by a detector within an element of
solid angle $d\Omega (\theta ,\varphi )$ in direction $(\theta ,\varphi )$
per unit time (to be designated $\delta \dot{N}$), and the flux of incident
particles, namely the number of particles that enter the sample per unit of
time and unit of area (to be designated $\Phi _{0}$). The latter is given by
%
\begin{equation}
\Phi _{0}=nv_{0}\;,
\end{equation}
where $n$ is the density of incident particles and $v_{0}$ their mean
velocity (e.g. velocity of light in the case of photons, the thermal
velocity in the case of thermalized neutrons). On the other hand we have
that
%
\begin{equation}
\delta \dot{N}=nV\sum_{\mathbf{p}_{1}^{\prime }\in d\Omega
}w_{\mathbf{p}_{0}\rightarrow \mathbf{p}_{1}^{\prime }}(\Delta E)\;,
\end{equation}
where $w$ is the probability per unit time that an incident particle
with momentum $\mathbf{p}_{0}$ makes a transition to a state of
momentum $\mathbf{p}_{1}^{\prime }$, the latter in directions
contained in the solid angle $ d\Omega (\theta ,\varphi )$ whose
axis we indicate by $\mathbf{p}_{1}$ (or $\hbar \mathbf{k}_{1}$);
$\Delta E$ is the transfer of energy in the scattering event, and
$nV$ is the number of particles and $V$ is the active volume of the
sample, i.e. the region involved in the process, for example the
region of focalization of the laser beam in the scattering of
photons.

However, since $d\Omega $ is small (fixed by the size of the
detector window) we can take all the contributions in the sum over
$\mathbf{p}_{1}$ as the same, i.e. to a good degree of
approximation equal to $w_{\mathbf{p}_{0}\rightarrow
\mathbf{p}_{1}}(\Delta E)$. Hence
%
\begin{equation}
d^{2}\sigma (\Delta E,\mathbf{q})=\frac{\delta \dot{N}}{\Phi
_{0}}=\frac{V}{ v_{0}}w_{\mathbf{p}_{0}\rightarrow
\mathbf{p}_{1}}(\Delta E)b (\mathbf{p}_{1},d\Omega )\;,
\end{equation}
where
%
\begin{equation}
b (\mathbf{p},d\Omega )=\sum_{\mathbf{p}_{1}^{\prime }\in d\Omega
}1\simeq \frac{V}{(2\pi \hbar )^{3}}p_{1}^{2}dp_{1}d\Omega
\end{equation}
is the number of states of the particles in the scattered beam, which are
entering the detector, and we recall that $\mathbf{p}=\hbar \mathbf{k}$ and
the sum is over the plane-wave state of wavevector $\mathbf{k}$.

Therefore, for calculating the differential cross section we need
to evaluate the transition probability per unit time, $w$. For
that purpose let us consider a system with Hamiltonian
$\Hat{H}_{\sigma}$, let us call $\Hat{H}_P$ the Hamiltonian of the
particles used in the experiment, and $\Hat{\mathcal{V}}$ the
interaction potential between the system and the particles, that
is
%
\begin{equation}
\Hat{H}=\Hat{H}_{\sigma}+\Hat{H}_{P}+\Hat{\mathcal{V}}\;.
\end{equation}

Let us introduce the notation $|\mu \rangle $ and $|\mathbf{p}\rangle $ for
the eigenfunctions of the system and the particles in the probe, i.e.
%
\begin{equation}
\Hat{H}_{\sigma }|\mu \rangle =E_{\mu }|\mu \rangle \;,
\end{equation}
%
\begin{equation}
\Hat{H}_{P}|\mathbf{p}\rangle =\hbar \omega
_{\mathbf{p}}|\mathbf{p}\rangle \;,
\end{equation}
where states $|\mathbf{p}\rangle $ are plane waves for the free particle
with momentum $\mathbf{p}$ in the incident and scattered beams. Moreover,
let
%
\begin{equation}
|\psi (t_{i})\rangle =|\Phi (t_{i})\rangle
|\mathbf{p}_{0}(t_{i})\rangle \label{TFReq109}
\end{equation}
be the wavefunction at the initial time $t_{i}$, that is, the initial
preparation of the system in the experiment.

As we have seen in previous sections, it is convenient to work in
the interaction representation, and then the wavefunction of the
system and probe at time $t$ is
%
\begin{equation}
|\psi(t)\rangle=U_0(t,t_i)U^{\prime}(t,t_i)|\psi(t_i)\rangle
\end{equation}
[cf. Eqs. (5) to (8)], with
%
\begin{equation}
i\hbar\frac{\partial}{\partial t}
U^{\prime}(t,t_i)=\tilde{\mathcal{V}}(t)U^{\prime}(t,t_i)\;,
\end{equation}
where $\tilde{\mathcal{V}}(t)$ is the potential in the interaction
representation, i.e. evolving with
$\Hat{H}_0=\Hat{H}_{\sigma}+\Hat{H}_p$, [cf. Eqs. (9) and (10)],
$U_0$ is given by
%
\begin{eqnarray}
U_{0}(t,t_{i}) &=&U_{\sigma }(t,t_{i})U_{P}(t,t_{i})  \notag \\
&=&e^{\frac{1}{i\hbar }(t-t_{i})\Hat{H}_{\sigma }}e^{\frac{1}{i\hbar
} (t-t_{i})\Hat{H}_{P}}\;,
\end{eqnarray}
and we recall that the iterated solution of Eq. (10) is given in Eq.
(11).

Moreover, if we define the function
%
\begin{equation}
|\tilde{\psi}(t)\rangle=U_0^{\dagger}(t,t_i)|\psi(t)\rangle
=U^{\prime}(t,t_i)|\psi(t_i)\rangle\;,
\end{equation}
we can easily verify that it satisfies the equation
%
\begin{equation}
i\hbar\frac{\partial}{\partial t}|\tilde{\psi}(t)\rangle
=\tilde{\mathcal{V}}(t)|\tilde{\psi}(t)\rangle\;,
\end{equation}
with $|\tilde{\psi}(t_i)\rangle=|\psi(t_i)\rangle$, which can be rewritten
as
%
\begin{equation}
|\tilde{\psi}(t)\rangle=|\psi(t_i)\rangle+ \frac{1}{i\hbar}\int_{t_i}^t
dt^{\prime}\,\tilde{\mathcal{V}}(t^{\prime}) |\tilde{\psi}%
(t^{\prime})\rangle\;,
\end{equation}
and then has the iterated solution
%
\begin{equation}
|\tilde{\psi}(t)\rangle=\Biggl[\, \sum_{n=0}^{\infty}
\frac{1}{(i\hbar)^n} \int_{t_i}^t dt_1\cdots\int_{t_i}^{t_{n-1}}
dt_n\,\tilde{\mathcal{V}}(t_1)\cdots\tilde{\mathcal{V}}(t_n)
\Biggr]|\psi(t_i)\rangle\;,
\end{equation}
and for later convenience in the calculations we notice that we can write
that
%
\begin{equation}
\tilde{\mathcal{V}}(t)|\tilde{\psi}(t)\rangle=\tilde{\varXi}(t)
|\psi(t_i)\rangle\;,
\end{equation}
with $\tilde{\varXi}$, called the scattering operator, satisfying the
integral equation
%
\begin{equation}
\tilde{\varXi}(t)=\tilde{\mathcal{V}}(t) \Biggl[1+ \frac{1}{i\hbar}
\int_{t_i}^t dt^{\prime}\,\tilde{\varXi}(t^{\prime})\Biggr]\;.
\end{equation}

It can be noticed that the right hand side of Eq. (128) provides the
effect of the perturbation $\mathcal{V}$ in all orders over the
initial nonperturbed wavefunction, equivalent to the effect in first
order over the interaction-representation function
$|\tilde{\psi}(t)\rangle$.

Let us now fix the scattering channel, i.e. we consider, as required
by Eq. (115), the scattering event with probe particles making
transitions between states of momentum $|\mathbf{p}_{0}\rangle $ and
$|\mathbf{p}_{1}\rangle $. According to the general theory of
Quantum Mechanics, the probability for this event at time $t$ is
given by
%
\begin{equation}
P_{\mathbf{p}_{0}\rightarrow \mathbf{p}_{1}}(t)=\sum_{\mu
}|\langle \mathbf{p }_{1},\mu |\psi (t)\rangle |^{2}\;,
\end{equation}
where the summation over all states $|\mu \rangle $ of the system will be
\textit{a posteriori} restricted by the selection rule involving
conservation of energy and momentum in the scattering events.

We can rewrite Eq. (131) as
%
\begin{eqnarray}
P_{\mathbf{p}_{0}\rightarrow \mathbf{p}_{1}}(t) &=&\sum_{\mu
}|\langle \mathbf{p}_{1},\mu |U_{0}(t,t_{i})U^{\prime
}(t,t_{i})|\psi (t_{i})\rangle
|^{2}  \notag \\
&=&\sum_{\mu }|\langle \mathbf{p}_{1},\mu
|U_{0}(t,t_{i})|\tilde{\psi} (t)\rangle |^{2}\;,
\end{eqnarray}
where we have used Eq. (125). On the other hand
%
\begin{equation}
\langle \mathbf{p},\mu |U_{0}(t,t_{i})=\langle \mathbf{p},\mu
|\exp \Biggl\{-\frac{1}{i\hbar }(t-t_{i})(E_{\mu }+\hbar \omega
_{\mathbf{p}})\Biggr\}\;,
\end{equation}
then the exponential gives a modulus 1 in Eq. (132), and we have
that
%
\begin{eqnarray}
P_{\mathbf{p}_{0}\rightarrow \mathbf{p}}(t) &=&\sum_{\mu }|\langle
\mathbf{p}
_{1},\mu |\psi (t)\rangle |^{2}  \notag \\
&=&\sum_{\mu }|\langle \mathbf{p}_{1},\mu |[1+\frac{1}{i\hbar }
\int_{t_{i}}^{t}dt^{\prime }\,\tilde{\varXi}(t^{\prime })]|\psi
(t_{i})\rangle |^{2}  \notag \\
&=&\sum_{\mu }|\langle \mathbf{p}_{1},\mu |\frac{1}{i\hbar }
\int_{t_{i}}^{t}dt^{\prime }\tilde{\varXi}(t^{\prime })|\Phi
(t_{i}),\mathbf{ p}_{0}\rangle |^{2}\;,
\end{eqnarray}
where we have used Eqs. (127) and (129), together with Eq. (121),
and the fact that the plane wave states $|\mathbf{p}_{0}\rangle $
and $|\mathbf{p}_{1}\rangle $ are orthonormal states.

Introducing
%
\begin{equation}
\tilde{\varXi}_{\mathbf{q}}(t)=\langle
\mathbf{p}_{1}|\tilde{\varXi}(t)| \mathbf{p}_{0}\rangle =\int
d^{3}r\,\frac{e^{\frac{1}{i\hbar }\mathbf{p}_{1}\cdot
\mathbf{r}}}{\sqrt{V}}\tilde{\varXi}(t)\frac{e^{-\frac{1}{i\hbar}
\mathbf{p}_{0}\cdot \mathbf{r}}}{\sqrt{V}}\;,
\end{equation}
with, we recall, $\hbar \mathbf{q}=\mathbf{p}_{0}-\mathbf{p}_{1}$, and using
that the squared modulus can be written as the product of the complex number
times its complex conjugate, we can write
%
\begin{eqnarray}
P_{\mathbf{p}_{0}\rightarrow \mathbf{p}_{1}}(t) &=&\frac{1}{\hbar
^{2}} \sum_{\mu }\langle \phi (t_{i})|\int_{-\infty }^{t}dt^{\prime
\prime }\tilde{ \varXi}_{\mathbf{q}}^{\dagger }(t^{\prime \prime
})|\mu \rangle \langle \mu |\int_{-\infty }^{t}dt^{\prime
}\,\tilde{\varXi}_{\mathbf{q}}(t^{\prime
})|\Phi (t_{i})\rangle   \notag \\
&=&\frac{1}{\hbar ^{2}}\int_{-\infty }^{t}dt^{\prime }\int_{-\infty
}^{t}dt^{\prime \prime
}Tr\bigl\{\tilde{\varXi}_{\mathbf{q}}^{\dagger }(t^{\prime \prime
})\tilde{\varXi}_{\mathbf{q}}(t^{\prime })\mathcal{P}
_{0}(t_{i})\bigr\}\;,
\end{eqnarray}
with $\mathcal{P}_{0}(t_{i})$ being the projection operator
(statistical operator for the pure state $|\Phi (t_{i})\rangle $ of
Eq. (121), and we have considered adiabatic application of the
perturbation in $t_{i}\rightarrow -\infty $), implying in that the
initial ultrafast transient is ignored.

So far we have a purely quantum-mechanical calculation, and we have
an expression depending on the initial preparation of the system as
characterized by the statistical operator for the pure state given
above. We need next to introduce the statistical average over the
mixed state, what is done averaging over the corresponding Gibbs
ensemble of all possible initial pure states compatible with the
thermodynamic condition of preparation of the system at time
$t_{i}$, that is, we do have have that
%
\begin{equation}
\langle P_{\mathbf{p}_{0}\rightarrow \mathbf{p}_{1}}(t)\rangle
=\frac{1}{ \hbar ^{2}}\int_{-\infty }^{t}dt^{\prime }\int_{-\infty
}^{t}dt^{\prime \prime
}Tr\bigl\{\tilde{\varXi}_{\mathbf{q}}^{\dagger }(t^{\prime \prime
}) \tilde{\varXi}_{\mathbf{q}}(t^{\prime })\varrho _{\varepsilon
}(t_{i})\times \varrho _{R}\bigr\}
\end{equation}
where $\varrho _{\varepsilon }(t_{i})\times \varrho
_{R}=\mathcal{R}_{\varepsilon }(t_{i})$ is the corresponding
statistical operator, involving $\varrho _{\varepsilon }$ of the
system in interaction with the thermal bath and $\varrho _{R}$ of
the thermal bath, which has been assumed to constantly remain in
equilibrium at temperature $T_{0}$.

The rate of transition probability $w$ in Eq. (116), is then
%
\begin{eqnarray}
w_{\mathbf{p}_{0}\rightarrow \mathbf{p}_{1}}(\Delta E|t)
&=&\frac{d}{dt}
\langle P_{\mathbf{p}_{0}\rightarrow \mathbf{p}_{1}}(t)\rangle   \notag \\
&=&\frac{1}{\hbar ^{2}}\int_{-\infty }^{t}dt^{\prime
}Tr\{\tilde{\varXi}_{ \mathbf{q}}^{\dagger
}(t)\tilde{\varXi}_{\mathbf{q}}(t^{\prime })\varrho
_{\varepsilon }(t_{i})\times \varrho _{R}\}+  \notag \\
&&\frac{1}{\hbar ^{2}}\int_{-\infty }^{t}dt^{\prime
}Tr\{\tilde{\varXi}_{ \mathbf{q}}^{\dagger }(t^{\prime
})\tilde{\varXi}_{\mathbf{q}}(t)\varrho _{\varepsilon }(t_{i})\times
\varrho _{R}\}\,\,.
\end{eqnarray}
Using that
%
\begin{equation}
Tr\{U_{0}^{\dagger }(t,t_{i})\tilde{\varXi}_{\mathbf{q}}^{\dagger
}U(t,t_{i})U_{0}^{\dagger }(t^{\prime
},t_{i})\tilde{\varXi}_{\mathbf{q} }U_{0}(t^{\prime
},t_{i})\mathcal{R}_{\varepsilon }(t_{i})\}=Tr\{\tilde{
\varXi}_{\mathbf{q}}^{\dagger }\tilde{\varXi}_{\mathbf{q}}(t^{\prime
}-t)\varrho _{\varepsilon }(t)\times\varrho _{R}\}\;,
\end{equation}
and that
%
\begin{eqnarray}
\tilde{\varXi}_{\mathbf{q}}(t^{\prime }-t) &=&\langle \mathbf{p}
_{1}|U_{P}^{\dagger }(t^{\prime }-t)U_{\sigma }^{\dagger }(t^{\prime
}-t) \tilde{\varXi}_{\mathbf{q}}U_{\sigma }(t^{\prime
}-t)U_{P}(t^{\prime }-t)|
\mathbf{p}_{0}\rangle   \notag \\
&=&e^{-\frac{1}{i\hbar }(t^{\prime }-t)\hbar (\omega _{\mathbf{p}
_{0}}-\omega
_{\mathbf{p}_{1}})}\tilde{\varXi_{\mathbf{q}}}(t^{\prime
}-t)_{\sigma }\;,
\end{eqnarray}
where $\tilde{\varXi}_{\mathbf{q}}(t^{\prime }-t)_{\sigma
}=U_{\sigma }^{\dagger }(t^{\prime
}-t,t_{i})\tilde{\varXi}_{\mathbf{q}}U_{\sigma }(t^{\prime
}-t,t_{i})$, we can write
%
\begin{eqnarray}
w_{\mathbf{p}_{0}\rightarrow \mathbf{p}_{1}}(\Delta E|t) &\equiv
&w(\mathbf{q
},\omega |t)  \notag \\
&=&\frac{1}{\hbar ^{2}}\int_{-\infty }^{t}dt^{\prime }\,e^{i\omega
(t^{\prime }-t)}Tr\{\tilde{\varXi}_{\mathbf{q}}^{\dagger
}\tilde{\varXi}_{ \mathbf{q}}(t^{\prime }-t)_{\sigma }\varrho
_{\varepsilon }(t)\times \varrho _{R}\}+\text{ c.c.}\;,
\end{eqnarray}
where $\omega =\omega _{\mathbf{p}_{0}}-\omega _{\mathbf{p}_{1}}$, and it
can be noticed that the statistical operator is given at time $t$, when a
measurement is performed.

If we consider the case of equilibrium, i.e. we take the canonical
distribution $\varrho _{c}$ instead of $\varrho _{\varepsilon }(t)$, and
taking into account that $\varrho _{c}$ and $H_{\sigma }$ commute, we find
that
%
\begin{equation}
w(\mathbf{q},\omega )_{\text{eq}}=\frac{1}{\hbar
^{2}}\int_{-\infty }^{\infty }d\tau ^{\prime }\,e^{-i\omega \tau
^{\prime }}Tr\{\tilde{\varXi}_{ \mathbf{q}}^{\dagger }(\tau
^{\prime })_{\sigma }\tilde{\varXi}_{\mathbf{q} }\varrho _{c}\}\;,
\end{equation}
which is the known temperature-dependent rate of transition
probability (see for example [50]).

In conclusion, the differential cross section is then given by
%
\begin{eqnarray}
\frac{d^{2}\sigma (\mathbf{q},\omega |t)}{d\omega \,d\Omega }
&=&\frac{V^{2} }{(2\pi \hbar )^{3}}\frac{g(\omega
)}{v_{0}}\frac{1}{\hbar ^{2}}\Biggl[ \int_{-\infty }^{t}dt^{\prime
}\,e^{i\omega (t^{\prime }-t)}Tr\{\tilde{\varXi
}_{\mathbf{q}}^{\dagger }\tilde{\varXi}_{\mathbf{q}}(t^{\prime
}-t)_{\sigma }
\mathcal{R}_{\varepsilon }(t)\}+  \notag \\
&&\int_{-\infty }^{t}dt^{\prime }\,e^{-i\omega (t^{\prime
}-t)}Tr\{\tilde{ \varXi}_{\mathbf{q}}^{\dagger }(t^{\prime
}-t)_{\sigma }\tilde{\varXi}_{ \mathbf{q}}\varrho _{\varepsilon
}(t)\times \varrho _{R}\}\Biggr]\;,
\end{eqnarray}
each term within the square bracket is the complex conjugate of the other
and then the quantity is real, as it should, and we have defined
%
\begin{equation}
p^{2}dp=g(\omega )\,d\omega \;,
\end{equation}
introducing the density of states $g(\omega )$ which follows in each case
once it is given the dispersion relation $\omega _{\mathbf{p}}$.

Moreover it is stressed the fact that, differently to the case when
the system is in equilibrium, in the nonequilibrium initial
preparation of the sample \emph{the scattering cross section is not
closed in itself, but it needs to be coupled to the set of kinetic
equations that describe the evolution of the out of equilibrium
system}, i.e., those that determine the statistical operator
$\mathcal{R}_{\varepsilon}(t)$. It can be noticed that this is a
question also present in the response function theory of the
previous sections.

As it has been noticed in previous Section, we can write
$\mathcal{R}_{\varepsilon }(t)=\varrho _{\varepsilon }(t)\times
\varrho _{R}$ and make use of the separation $\varrho _{\varepsilon
}(t)=\bar{\varrho}(t)+\varrho _{\varepsilon }^{\prime }(t)$ to
obtain that
%
\begin{equation}
d^{2}\sigma (\mathbf{q},\omega |t)=d^{2}\bar{\sigma}(\mathbf{q},\omega
|t)+d^{2}\bar{\sigma}_{\varepsilon }^{\prime }(\mathbf{q},\omega |t)\,,
\end{equation}
that is, the contribution $d^{2}\bar{\sigma}$ where the trace is
taken with $\bar{\varrho}$ and $d^{2}\sigma_{\varepsilon }^{\prime
}$ with the trace taken with $\varrho _{\varepsilon }^{\prime }$.

Let us consider the case of time- and space- resolved scattering,
that is the detector in Fig. 2 collects the scattered particles
arriving from an element of volume $\Delta V(\mathbf{r})$ around
position $\mathbf{r}$ in the sample. For simplicity we take the
first-order scattering consisting that in Eq. (142) we take of the
scattering operator $\Xi $ of Eq. (130) the first contribution
$\mathcal{\bar{V}}$, and for the latter we write
%
\begin{equation}
\mathcal{\hat{V}}=\sum\limits_{\mu =1}^{N^{\prime
}}\sum\limits_{j=1}^{N}\upsilon (\mathbf{R}_{\mu }-\mathbf{r}_{j})\,,
\end{equation}
where $\mathbf{r}_{j}$ is the position of \emph{j}-th particle in
the system and $\mathbf{R}_{\mu }$ the position of the $\mu $-th
particle in the incident beam.

Therefore, we have that
%
\begin{equation}
\mathcal{\hat{V}}_{\mathbf{q}}=\langle
\mathbf{p}_{0}|\mathcal{\hat{V}}| \mathbf{p}_{1}\rangle
=n_{b}\upsilon (\mathbf{q})\sum\limits_{j=1}^{N}e^{i
\mathbf{q}\cdot \mathbf{r}_{j}}\,,
\end{equation}
where we have introduced the Fourier amplitude
%
\begin{equation}
\upsilon (\mathbf{q})=\int d^{3}b \upsilon (b)e^{i\mathbf{q}\cdot
\mathbf{b}}\,,
\end{equation}
with $\mathbf{b} =\mathbf{r}-\mathbf{r}_{j}$, and we recall that
$\hbar \mathbf{q}=\mathbf{p}_{0}-\mathbf{p}_{1}$, and $n_{b}$ is the
density of particles in the beam.

Retaining in Eq. (143) the contribution in first order in
$\mathcal{V}$ only, it follows that
%
\begin{equation}
\frac{d^{2}\sigma (\mathbf{q},\omega |t)}{d\omega d\Omega
}=\frac{V^{2}}{ (2\pi \hbar )^{3}}\frac{g(\omega )}{\hbar
^{2}v_{0}}n_{b}^{2}|\upsilon (
\mathbf{q})|^{2}S_{nn}(\mathbf{q},\omega |t)\,,
\end{equation}
where
%
\begin{equation}
S_{nn}(\mathbf{q},\omega |t)=\sum\limits_{j,l}\int_{-\infty
}^{t}dt^{\prime }\,e^{i\omega (t^{\prime
}-t)}Tr\{e^{-i\mathbf{q}\cdot \lbrack \mathbf{r} _{j}(t^{\prime
}-t)-\mathbf{r}_{l}]}\varrho _{\varepsilon }(t)\times\varrho _{R}\}+
\text{c.c.}\;,
\end{equation}

We introduce now the density operator
%
\begin{equation}
\hat{n}(\mathbf{r},\tau )=\sum\limits_{j}\delta
(\mathbf{r}-\mathbf{r}_{j}(t))
\end{equation}
and then we can rewrite the correlation function of Eq. (150) as
%
\begin{equation}
S_{nn}(\mathbf{q},\omega |t)=\int d^{3}r\int d^{3}r^{\prime
}\int_{-\infty }^{t}dt^{\prime }\,e^{i\omega (t^{\prime
}-t)}e^{-i\mathbf{q}\cdot (\mathbf{r }-\mathbf{r}^{\prime
})}Tr\{\hat{n}^{\dag }(\mathbf{r},t^{\prime }-t)\hat{n}(
\mathbf{r}^{\prime })\varrho _{\varepsilon }(t)\times\varrho
_{R}\}+\text{c.c.}\;,
\end{equation}
where the integrations in space run over the active volume of the
sample (region of concentration of the particle beam), or in the
case of a space-resolved experiment over $\Delta V(\mathbf{r})$ and
then we do have the time- and space- resolved spectrum
%
\begin{equation}
\frac{d^{2}\sigma (\mathbf{r};\mathbf{q},\omega |t)}{d\omega d\Omega
} =\Delta V(\mathbf{r})\int d^{3}r^{\prime }\int_{-\infty
}^{t}dt^{\prime }\,e^{i\omega (t^{\prime }-t)}e^{-i\mathbf{q}\cdot
(\mathbf{r}-\mathbf{r} ^{\prime })}Tr\{\hat{n}^{\dag
}(\mathbf{r},t^{\prime }-t)\hat{n}(\mathbf{r} ^{\prime })\varrho
_{\varepsilon }(t)\times \varrho _{R}\}+\text{c.c.}\,,
\end{equation}

In the case of an experiment in photoluminescence (recombination of
photoexcited electrons and holes) the potential has the form
%
\begin{equation}
\mathcal{V}=\sum\limits_{j}\mathbf{A}(\mathbf{r}_{j},t)\cdot
\mathbf{p}_{j}\,,
\end{equation}
and then, after neglecting the photon momentum (dipolar approximation), the
luminescence spectrum is given by
%
\begin{equation}
P_{L}(\mathbf{r};\omega |t)\sim
\sum\limits_{\mathbf{k}}\bar{f}_{\mathbf{k}
}^{\,e}(\mathbf{r},t)\bar{f}_{\mathbf{k}}^{\,h}(\mathbf{r},t)\delta
(\hbar k^{2}/2m_{x}+E_{G}-\hbar \omega )\,,
\end{equation}
in arbitrary units, where we have introduced a local approximation
and it has been used the effective mass approximation for electrons
($e$) and holes ($h$), $m_{x}^{-1}=m_{e}^{-1}+m_{h}^{-1}$ is the
excitonic mass, $E_{G}$ is the energy gap, and
$\bar{f}_{\mathbf{k}}^{\,e(h)}(\mathbf{r},t)$ are the populations in
state $\mathbf{k}$, in position $\mathbf{r}$ and at time $t$ of
electrons (holes) given by
%
\begin{equation}
\bar{f}_{\mathbf{k}}^{\,e(h)}(\mathbf{r},t)=[1+\exp \{\beta ^{\ast
}(\mathbf{r},t)[\hbar ^{2}k^{2}/2m_{e(h)}-\mu _{e(h)}^{\ast
}(\mathbf{r},t)]^{-1}\}\,,
\end{equation}
where $\beta ^{\ast }(\mathbf{r},t)$ is the reciprocal of the field
of nonequilibrium temperature and $\mu _{e(h)}^{\ast
}(\mathbf{r},t)$ the quasi-chemical potential.


\section{Illustrative Examples}

\subsection{Experiments in ultra-fast-laser spectroscopy.}

Pump-probe experiments in the field of ultrafast laser spectroscopy,
devoted to the study of the nonequilibrium photoinjected plasma in
semiconductors, have been extensively used in recent decades, and
have been accompanied by a number of theoretical analysis [58-64].
In this type of experiments the system is, as a general rule, driven
far away from equilibrium and consequently, its theoretical
description falls into the realm of the thermodynamics of
irreversible processes in far-from-equilibrium systems, and the
accompanying kinetic and statistical theories, and a particularly
appropriate approach is the \textsc{nesef} described here. The
theory presented in the previous Section has been used to derive in
detail a response function theory for the study of ultrafast optical
properties in the photoinjected plasma in semiconductors.
Particularly, one needs to derive the frequency- and wave
number-dependent dielectric function in arbitrary nonequilibrium
conditions, because it is the quantity which contains all the
information related to the optical properties of the system (it
provides the absorption coefficient, the reflectivity coefficient,
the Raman scattering cross section, and so on). This is described
below, and moreover, we describe the application of the results to
the study of a particular type of experiment, namely the
time-resolved reflectivity changes in GaAs and other materials
[51-54] where signal changes in the reflectivity, $\Delta R/R$, of
the order of $10^{-7}$ are detected, and a distinct oscillation of
the signal in real time is observed. In Fig. 3 are reproduced
time-resolved reflectivity spectra, and in the upper right inset is
shown the part corresponding to the observed oscillation, as
reported by Cho \textit{et al.} [51]

%
\begin{figure}[h]
\center
\includegraphics[width=10cm]{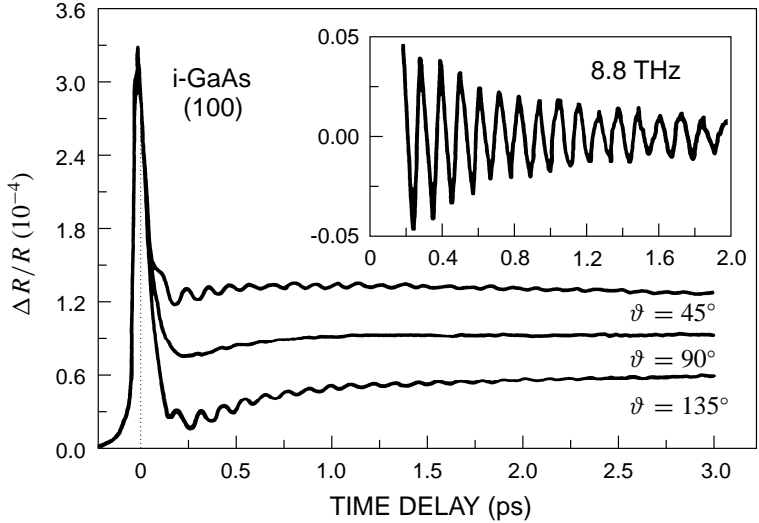}
\caption{Reproduction of the time-resolved reflectivity changes in
GaAs, as reported by Cho, K\"{u}tt, and Kurz in Ref. [51].}
\end{figure}

Such phenomenon has been attributed to the generation of coherent
lattice vibrations, and several theoretical approaches have been
reported [54-56]. A clear description, on phenomenological bases,
which captures the essential physics of the problem, is reported in
Ref. [55], and in Ref. [57] is presented an analysis based on
\textsc{nesef}, where the different physical aspects of the problem
are discussed. It is evidenced that the oscillatory effect is
provided by the displacive excitation of the polar lattice
vibrations, arising out of the coupling of the carrier-charge
density and polar modes, and its decay is mainly governed by the
cooling down of the carriers. We briefly describe the experiment and
all the pertaining \textsc{nesef}-based calculations in
continuation.

Let us consider a direct-gap polar semiconductor in a pump-probe experiment.
We recall that the exciting intense laser pulse produces the so-called
highly excited plasma in semiconductors, namely, electron-hole pairs in the
metallic side of Mott transition (that is, they are itinerant carriers, and
we recall that this requires concentrations of these photoinjected
quasi-particles of order of $10^{16}$~cm$^{-3}$ and up), which compose a
two-component Fermi fluid, moving in the lattice background. It constitutes
a highly nonequilibrated system where the photoexcited carriers rapidly
redistribute their energy in excess of equilibrium via, mainly, the strong
long-range Coulomb interaction (pico- to subpico- second scale), followed by
the transfer of energy to the phonon field (predominantly to the optical
phonons, and preferentially to the \textsc{lo} phonons via Fr\"ohlich
interaction), and finally via acoustic phonons to the external thermal
reservoir. Along the process the carrier density diminishes in recombination
processes (nanosecond time scale) and through ambipolar diffusion out of the
active volume of the sample (ten-fold picosecond time scale).

Moreover, a probe interacting weakly with the \textsc{heps} is used
to obtain an optical response, namely the reflectivity of the
incoming laser photons with frequency $\omega $ and wave vector
$\mathbf{Q}$. From the theoretical point of view, such measurement
is to be analyzed in terms of, as already noted, the all important
and inevitable use of correlation functions in response function
theory. The usual application in normal probe experiments performed
on a system initially in equilibrium had a long history of success,
and a practical and elegant treatment is based on the method of the
double-time (equilibrium) thermodynamic Green functions [45,46]. In
the present case of a pump-probe experiment we need to resort to a
theory of such type but applied to a system whose macroscopic state
is in nonequilibrium conditions and evolving in time as a result of
the dissipative processes that are developing while the sample is
probed. This is, resorting to the theory of previous section, and we
recall that the response function theory for nonequilibrium systems
needs be coupled to the kinetic theory that describes the evolution
of the nonequilibrium state of the system. We resort here to such
theory for the study of the reflectivity experiments of Ref. [51].

The time-dependent (because it keeps changing along with the
evolution of the macrostate of the nonequilibrated system)
reflectivity $R(\omega , \mathbf{Q}|t)$ is related to the index of
refraction $\eta (\omega ,\mathbf{Q }|t)+i\kappa (\omega
,\mathbf{Q}|t)$ through the well-known expression
%
\begin{equation}
R(\omega ,\mathbf{Q}|t)= \frac{\left[ \eta (\omega
,\mathbf{Q}|t)-1\right] ^{2}+\left[ \kappa (\omega
,\mathbf{Q}|t)\right] ^{2}}{\left[ \eta (\omega
,\mathbf{Q}|t)+1\right] ^{2}+ \left[ \kappa (\omega
,\mathbf{Q}|t)\right] ^{2}}\;,
\end{equation}
and the refraction index is related to the time-evolving frequency- and wave
vector-dependent dielectric function by
\begin{eqnarray}
\epsilon (\omega ,\mathbf{Q}|t) &=&\epsilon ^{\prime }(\omega
,\mathbf{Q}
|t)+i\epsilon ^{\prime \prime }(\omega ,\mathbf{Q}|t)  \notag \\
&=&\left[ \eta (\omega ,\mathbf{Q}|t)+i\kappa (\omega
,\mathbf{Q}|t)\right] ^{2}\;,
\end{eqnarray}
where $\eta $ and $\epsilon ^{\prime }$, and $\kappa $ and $\epsilon
^{\prime \prime }$, are the real and imaginary parts of the refraction index
and of the dielectric function respectively.

We call the attention to the fact that the dielectric function depends on
the frequency and the wave vector of the radiation involved, and $t$ stands
for the time when a measurement is performed. Once again we stress that this
dependence on time is, of course, the result that the macroscopic state of
the non-equilibrated plasma is evolving in time as the experiment is
performed.

Therefore it is our task to calculate this dielectric function in the
nonequilibrium state of the system. First, we note that according to Maxwell
equations in material media (that is, Maxwell equations now averaged over
the nonequilibrium statistical ensemble) we have that
%
\begin{equation}
\epsilon ^{-1}(\omega ,\mathbf{Q}|t)-1=\frac{n(\omega
,\mathbf{Q}|t)}{r(\omega ,\mathbf{Q})}\;,
\end{equation}
where $r(\omega ,\mathbf{Q})$ is the amplitude of a probe charge
density with frequency $\omega $ and wave vector $\mathbf{Q}$, and
$n(\omega ,\mathbf{Q}|t)$ is the induced polarization-charge density
of carriers and lattice in the media. As shown before the latter can
be calculated resorting to the response function theory for systems
far from equilibrium (the case is quite similar to the calculation
of the time-resolved Raman scattering cross section [48]), and
obtained in terms of the nonequilibrium-thermodynamic Green
functions, as we proceed to describe.

Using the formalism we have presented to obtain $\epsilon (\omega
,\mathbf{Q}|t)$, [cf. Eqs. (98)] it follows that
%
\begin{equation}
\epsilon ^{-1}(\omega ,\mathbf{Q})-1=V(\mathbf{Q})\,[G_{cc}(\omega
,\mathbf{Q })+G_{ci}(\omega ,\mathbf{Q})+G_{ic}(\omega
,\mathbf{Q})+G_{ii}(\omega , \mathbf{Q})]\;,
\end{equation}
giving the reciprocal of the dielectric function in terms of Green functions
given by
%
\begin{equation}
G_{cc}(\omega ,\mathbf{Q})=\langle \!\langle
\hat{n}_{c}(\mathbf{Q})\,;\, \hat{n}_{c}^{\dagger
}(\mathbf{Q})|\,\omega ;t\rangle \!\rangle \;,
\end{equation}
%
\begin{equation}
G_{ci}(\omega ,\mathbf{Q})=\langle \!\langle
\hat{n}_{c}(\mathbf{Q})\,;\, \hat{n}_{i}^{\dagger
}(\mathbf{Q})|\,\omega ;t\rangle \!\rangle \;,
\end{equation}
%
\begin{equation}
G_{ic}(\omega ,\mathbf{Q})=\langle \!\langle
\hat{n}_{i}(\mathbf{Q})\,;\, \hat{n}_{c}^{\dagger
}(\mathbf{Q})|\,\omega ;t\rangle \!\rangle \;,
\end{equation}
%
\begin{equation}
G_{ii}(\omega ,\mathbf{Q})=\langle \!\langle
\hat{n}_{i}(\mathbf{Q})\,;\, \hat{n}_{i}^{\dagger
}(\mathbf{Q})|\,\omega ;t\rangle \!\rangle \;,
\end{equation}
where $V(\mathbf{Q})=4\pi ne^{2}/V\varepsilon _{0}Q^{2}$ is the
matrix element of the Coulomb potential in plane-wave states and
$\hat{n}_{c}(\mathbf{Q})$, and $\hat{n}_{i}(\mathbf{Q})$, refer to
the $\mathbf{Q}$-wave vector Fourier transform of the operators
for the densities of charge of carriers and the polarization
charge of longitudinal optical phonons respectively.

But, the expression we obtain is, as already noticed, depending on
the evolving nonequilibrium macroscopic state of the system, a fact
embedded in the expressions for the time-dependent distribution
functions of the carrier and phonon states. Therefore, they are to
be derived within the kinetic theory in \textsc{nesef}, and the
first and fundamental step is the choice of the set of variables
deemed appropriate for the description of the macroscopic state of
the system. In this case a first set of variables needs be the one
composed of the carriers' density and energy, and the phonon
population functions, together with the set of associated Lagrange
multipliers that, as we have seen, can be interpreted as a
reciprocal quasitemperature and quasi-chemical potentials of
carriers, and reciprocal quasitemperatures of phonons, one for each
mode [42,58,59]. But in the situation we are considering we need to
add, on the basis of the information provided by the experiment, the
amplitudes of the \textsc{lo}-lattice vibrations and the carrier
charge density; the former because it is clearly present in the
experimental data (the oscillation in the reflectivity) and the
latter because of the \textsc{lo}-phonon-plasma coupling clearly
present in Raman scattering experiments [60,61]). Consequently the
chosen basic set of dynamical quantities is
%
\begin{equation}
\{\Hat{H}_{c},\Hat{N}_{e},\Hat{N}_{h},\hat{n}_{\mathbf{kp}}^{e},\hat{n}_{
\mathbf{kp}}^{h},\hat{\nu}_{\mathbf{q}},a_{\mathbf{q}},a_{\mathbf{q}
}^{\dagger },H_{B}\}\;,
\end{equation}
where
%
\begin{equation}
\Hat{H}_{c}=\sum_{\mathbf{k}}\left[ \varepsilon
_{\mathbf{k}}^{e}\,c_{ \mathbf{k}}^{\dagger
}\,c_{\mathbf{k}}+\varepsilon _{\mathbf{k}}^{h}\,h_{-
\mathbf{k}}^{\dagger }\,h_{-\mathbf{k}}\right] \;,
\end{equation}
%
\begin{equation}
\hat{\nu}_{\mathbf{q}}=a_{\mathbf{q}}^{\dagger }\,a_{\mathbf{q}}\;,
\end{equation}
%
\begin{equation}
\Hat{N}_{e}=\sum_{\mathbf{k}}c_{\mathbf{k}}^{\dagger }c_{\mathbf{k}
}\,\,,\qquad \Hat{N}_{h}=\sum_{\mathbf{k}}h_{-\mathbf{k}}^{\dagger
}\,h_{- \mathbf{k}}\,\,,
\end{equation}
%
\begin{equation}
\hat{n}_{\mathbf{kp}}^{e}=c_{\mathbf{k}+\mathbf{p}}^{\dagger
}c_{\mathbf{k} }\,\,,\qquad
\hat{n}_{\mathbf{kp}}^{h}=h_{-\mathbf{k}-\mathbf{p}}h_{-\mathbf{
k}}^{\dagger }\,\,,
\end{equation}
with $c$ ($c^{\dagger }$), $h$ ($h^{\dagger }$), and $a$
($a^{\dagger }$) being as usual annihilation (creation) operators in
electron, hole, and \textsc{lo}-phonon states respectively
($\mathbf{k},\mathbf{p},\mathbf{q}$ run over the Brillouin zone).
Moreover, the effective mass approximation is used and Coulomb
interaction is dealt with in the random phase approximation, and
then \hbox{$\epsilon_{\mathbf{k}}^{e}=E_{G}+\hbar
^{2}|\mathbf{k}|^{2}/2m_{e}$} and
\hbox{$\epsilon_{\mathbf{k}}^{h}=\hbar^{2}|
\mathbf{k}|^{2}/2m_{h}$}. Finally $H_{B}$ is the Hamiltonian of the
lattice vibrations different from the \textsc{lo} one. We write for
the \textsc{nesef}-nonequilibrium thermodynamic variables associated
to the quantities of Eq. (165)
%
\begin{equation}
\{\beta _{c}(t),\;-\beta _{c}(t)\mu _{e}^{\ast }(t),\;-\beta _{c}(t)\mu
_{h}^{\ast }(t),\;F_{\mathbf{kp}}^{e}(t),\; \\
F_{\mathbf{kp}}^{h}(t),\;\hbar \omega _{\mathbf{q}}\beta
_{\mathbf{q}}(t),\;\varphi _{\mathbf{q}}(t),\varphi
_{\mathbf{q}}^{\ast }(t),\;\beta _{0}\}\;,
\end{equation}
respectively, where $\mu _{e}^{\ast }$ and $\mu _{h}^{\ast }$ are
the quasi-chemical potentials for electrons and for holes; we write
\hbox{$\beta_{c}(t)=1/k_{B}T_{c}^{\ast}(t)$} introducing the
carriers' quasitemperature $T_{c}^{\ast }$;
\hbox{$\beta_{\mathbf{q}}(t)=1/k_{B}T_{\mathbf{q}}^{\ast}(t)$}
introducing the \textsc{lo}-phonon quasitemperature per mode
($\omega _{\mathbf{q}}$ is the dispersion relation), $\beta
_{0}=1\,/\,k_{B}\,T_{0}$ with $T_{0}$ being the temperature of the
thermal reservoir. We indicate the corresponding macrovariables,
that is, those which define the nonequilibrium thermodynamic Gibbs
space as
%
\begin{equation}
\{E_{c}(t),\;n(t),\;n(t),\;n_{\mathbf{kp}}^{e}(t),\;n_{\mathbf{kp}
}^{h}(t),\nu _{\mathbf{q}}(t),\;\langle a_{_{\mathbf{q}}}|t\rangle
,\;\langle a_{\mathbf{q}}^{\dagger }|t\rangle =\langle
a_{\mathbf{q} }|t\rangle ^{\ast },\;E_{B}\}\;,
\end{equation}
which are the statistical average of the quantities of Eq. (165),
that is
%
\begin{equation}
E_{c}(t) =Tr\bigl\{\Hat{H}_{c}\,\varrho _{\epsilon }(t)\times
\varrho _{R} \bigr\}\;,
\end{equation}
%
\begin{equation}
n(t) =Tr\bigl\{\Hat{N}_{e(h)}\,\varrho _{\epsilon }(t)\times \varrho
_{R} \bigr\}\;,
\end{equation}
and so on, where $\varrho _{R}$ is the stationary statistical distribution
of the reservoir and $n(t)$ is the carrier density, which is equal for
electrons and for holes since they are produced in pairs in the intrinsic
semiconductor. The volume of the active region of the sample (where the
laser beam is focused) is taken equal to $1$ for simplicity.

Next step is to derive the equation of evolution for the basic
variables that characterize the nonequilibrium macroscopic state of
the system, and from them the evolution of the nonequilibrium
thermodynamic variables. This is done according to the generalized
\textsc{nesef}-based nonlinear quantum transport theory already
described, but in the second-order approximation in relaxation
theory. This is an approximation which retains only two-body
collisions but with memory being neglected, consisting in the
Markovian limit of the theory. It is sometimes referred to as the
quasi-linear approximation in relaxation theory [43,44], a name we
avoid because of the misleading word linear which refers to the
lowest order in dissipation, however the equations are highly
nonlinear.

The \textsc{nesef}-auxiliary (\textquotedblleft instantaneously
frozen\textquotedblright ) statistical operator is in the present
case given, in terms of the variables of Eq. (165) and the
nonequilibrium thermodynamic variables of Eq. (171), by
%
\begin{eqnarray}
\bar{\varrho}(t,0) &=&\exp \Bigl\{-\phi (t)-\beta _{c}^{\ast
}(t)[\Hat{H} _{c}-\mu _{e}^{\ast }(t)\,\Hat{N}_{e}-\mu _{h}^{\ast
}(t)\,\Hat{N}_{h}]-
\notag \\
&&\sum_{\mathbf{kp}}[F_{\mathbf{kp}}^{e}(t)\,\hat{n}_{\mathbf{kp}}^{e}(t)+F_{
\mathbf{kp}}^{h}(t)\,\hat{n}_{\mathbf{kp}}^{h}(t)]-  \notag \\
&&\sum_{\mathbf{q}}[\beta _{\mathbf{q}}^{\ast }(t)\,\hbar \,\omega
_{\mathbf{ q}}\,\hat{\nu}_{\mathbf{q}}+\varphi
_{\mathbf{q}}(t)\,a_{\mathbf{q}}+\varphi _{\mathbf{q}}^{\ast
}(t)\,a_{\mathbf{q}}^{\dagger }]-\beta _{0}\,H_{B}\Bigr\} \;,
\end{eqnarray}
where $\phi (t)$ ensures the normalization of $\bar{\varrho}(t,0)$.

Using such statistical operator the Green functions that define the
dielectric function [cf. Eq. (160)] can be calculated. This is an
arduous task, and in the process it is necessary to evaluate the
occupation functions
%
\begin{equation}
f_{\mathbf{k}}(t)=Tr\{c_{\mathbf{k}}c_{\mathbf{k}}\varrho _{\varepsilon
}(t)\}\;,
\end{equation}
which is dependent on the variables of Eq. (171). The
(nonequilibrium) carrier quasitemperature $T_{c}^{\ast }$ is
obtained, and its evolution in time shown in Fig. 4.

%
\begin{figure}[h]
\center
\includegraphics[width=10cm]{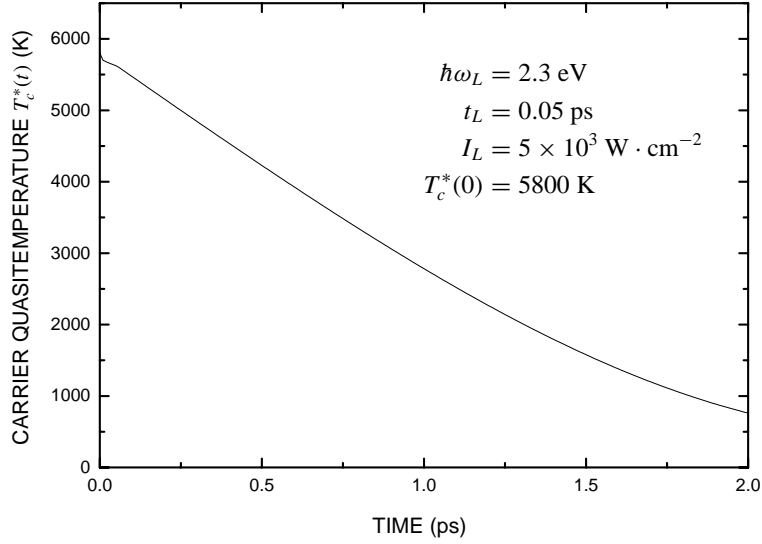}
\caption{Evolution of the carrier's quasitemperature, calculated in the
conditions of the experiment in the caption to Fig. 3.}
\end{figure}

Finally, in Fig. 5, leaving only as an adjustable parameter the amplitude --
which is fixed fitting the first maximum --, is shown the calculated
modulation effect which is compared with the experimental result (we have
only placed the positions of maximum and minimum amplitude taken from the
experimental data, which are indicated by the full squares).

%
\begin{figure}[h]
\center
\includegraphics[width=10cm]{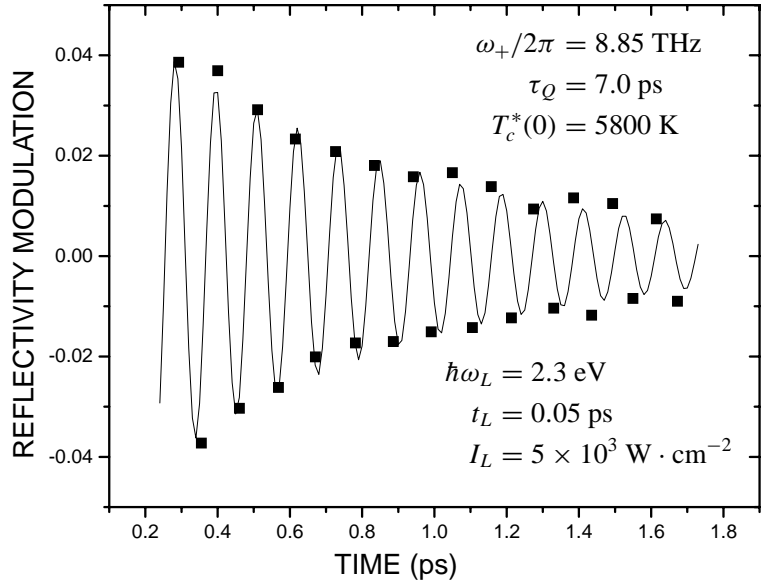}
\caption{The theoretically evaluated modulation of the time-resolved
reflectivity in the conditions of Ref. [51], compared with the
experimental data. For simplicity we have drawn only the positions
of the maxima and minima of the figure in the inset of Fig. 3.}
\end{figure}

This demonstrates the reason of the presence of the observed
modulating phenomenon in the reflectivity spectra, occurring with
the frequency of the near zone center \textsc{lo}-phonon (more
precisely the one of the upper $L_{+}$ hybrid mode [61]) with wave
vector $\mathbf{Q}$, the one of the photon in the laser radiation
field. The amplitude of the modulation is determined by the
amplitude of the laser-radiation-driven carrier charge density which
is coupled to the optical vibration, and then an open parameter in
the theory to be fixed by the experimental observation. This study
has provided, as shown, a good illustration of the full use of the
formalism of \textsc{nesef}, with an application to a quite
interesting experiment and where, we recall, the observed signal
associated to the modulation is seven orders of magnitude smaller
than the main signal on which is superimposed.

\subsection{Charge Transport in Doped Semiconductors}

NESEF is particularly appropriate do describe the transient and
steady state of semiconductors in the presence of intermediate to
strong electric fields (say tens to hundreds of kV/cm), fields
which drive the system far-from equilibrium. The question has
large technological interest because of the presence of such
situation in the integrated circuits of electronic and
optoelectronic devices.

Let us consider a n-doped direct-gap polar semiconductor, in
condition such that the extra electrons act as mobile carriers in
the conduction band. We use the effective-mass approximation, and
therefore parabolic band; this implies that in explicit
applications it needs be controlled the fact that there exists an
upper limiting value for the electric field strength, such that
below this limit we are working with field intensities for which
intervalley scattering can be neglected. The Hamiltonian of the
system is
%
\begin{equation}
\hat{H}=\hat{H}_{0}+\hat{H}_{1}+\hat{H}_{AN}+\hat{H}_{CF}+\hat{W}\,,
\end{equation}
where
%
\begin{equation}
\hat{H}_{0}=\sum\limits_{\mathbf{k}}(\hbar ^{2}k^{2}/2m_{e}^{\ast
})c_{ \mathbf{k}}^{\dagger }c_{\mathbf{k}}\text{
}+\sum\limits_{\mathbf{q},\gamma }\hbar \omega _{\mathbf{q},\gamma
}(b_{\mathbf{q}_{\gamma }}^{\dagger }b_{ \mathbf{q}_{\gamma
}}+1/2)\,,
\end{equation}
is the Hamiltonian of free electrons and phonons in branches
$\gamma =$ \textsc{lo},\textsc{ac}, and
%
\begin{equation}
\hat{H}_{1}=\sum\limits_{\mathbf{k},\mathbf{q},\gamma ,\sigma
}[M_{\gamma }^{\sigma }(\mathbf{q})b_{\mathbf{q}_{\gamma
}}c_{\mathbf{k}+\mathbf{q} _{\gamma }}^{\dagger
}c_{\mathbf{k}}+M_{\gamma }^{\sigma \ast }(\mathbf{q}
)b_{\mathbf{q}_{\gamma }}^{\dagger }c_{\mathbf{k}}^{\dagger
}c_{\mathbf{k}+ \mathbf{q}_{\gamma }}\,]\,,
\end{equation}
is the interaction Hamiltonian between them. In these equations
$c(c^{\dagger })$ and $b(b^{\dagger })$ are annihilation
(creation) operators in electron states $|\mathbf{k}>$, and of
phonons in mode $|\mathbf{q}_{\gamma }>$ and branches $\gamma
=$\textsc{lo},\textsc{ac} (for longitudinal optical and acoustical
ones respectively; \textsc{to}-phonons are ignored once they do
not interact with the electrons in the conduction band). Quantity
M\symbol{126}(q) is the matrix element of the interaction between
carriers and $\gamma $-type phonons, with superscript $\sigma$
indicating the kind of interaction (polar, deformation potential,
piezoelectric). Moreover, $\hat{H}_{AN}$ stands for the anharmonic
interaction in the phonon system, and
%
\begin{equation}
\hat{H}_{CF}=-\sum\limits_{i}e\mathbf{E}\cdot \mathbf{r}_{i},
\end{equation}
is the interaction of the electrons (with charge $-e$ and positions
$\mathbf{r}_{i}$) with an electric field $\mathbf{F}$ of intensity
$\mathcal{E}$. The interaction of the system with an external
reservoir is taken care of by $\hat{W}$ in Eq. (35); the reservoir
is taken as an ideal one -- what is satisfactory in most cases --
and then has its macroscopic (thermodynamic) state characterized by
a canonical statistical distribution with temperature $T_{0}$.

Consider now the nonequilibrium thermodynamic state of the system:
the presence of the electric field changes the energy of the
electrons (they acquire energy in excess of equilibrium), and
these carriers keep transferring this excess to the lattice and
from the lattice to the thermal reservoir, and an electrical
current (flux of electrons) follows. Thus, we need to choose as
basic variables
%
\begin{equation}
\bigl\{E_{e}(t),N_{e}(t),\mathbf{P}_{e}(t),E_{LO}(t),E_{AC}(t),E_{R}\bigr\},
\end{equation}
that is, respectively, the energy, number, and linear momentum of
the carriers, the energies of the \textsc{lo} and \textsc{ac}
phonons, and the energy of the reservoir; the latter is constant
in time for being considered as an ideal one. The corresponding
dynamical quantities are
%
\begin{equation}
\bigl\{\hat{H}_{e},\hat{N}_{e},\mathbf{\hat{P}}_{e},\hat{H}_{LO},\hat{H}
_{AC},\hat{H}_{R}\bigr\},
\end{equation}
i.e. the Hermitian operators for the partial Hamiltonians, the
electron number and the linear momentum. We noticed that the above
choice implies in disregarding electro-thermal effects, whose
inclusion would require to introduce the flux of energy (heat
current) of the carriers; it has a minor influence on the results
to be reported.

According to the nonequilibrium statistical ensemble formalism
described in section 2 of the preceding article, the nonequilibrium
thermodynamic state of the system, in an alternative description to
the one provided by the variables of Eq. (181), can be completely
characterized by a set of intensive nonequilibrium thermodynamic
variables (Lagrange multipliers that the variational construction of
the formalism provides), namely
%
\begin{equation}
\bigl\{F_{e}(t),F_{ne}(t),\mathbf{F}_{e}(t),F_{LO}(t),F_{AC}(t),\beta
_{0}\} \bigr\},
\end{equation}
The variables in this Eq. (182) are present in the auxiliary
statistical operator which the formalism introduces, in this case
given by
%
\begin{eqnarray}
\bar{\varrho}(t,0) &=&\exp \bigl\{-\phi
(t)-F_{ne}(t)\Hat{N}_{e}-F_{e}(t)
\Hat{H}_{e}-  \notag \\
&&-\mathbf{F}_{e}(t)\cdot \mathbf{\Hat{P}}_{e}-F_{LO}(t)\hat{H}
_{LO}-F_{AC}(t)\hat{H}_{AC}]\times \varrho _{R}\bigr\},
\end{eqnarray}
where $\varrho _{R}$ is the canonical distribution of the reservoir
at temperature $T_{0}$. We recall that the operator of Eq. (183) is
not the statistical operator describing the macroscopic state of the
system, which is a superoperator of this one, and $\phi (t)$
(playing the role of a logarithm of a nonequilibrium partition
function) ensures the normalization of $\bar{\varrho}(t,0)$ (see
Appendix A).

The intensive nonequilibrium thermodynamic variables of Eq. (182)
are usually redefined as
%
\begin{equation}
F_{e}(t)=\beta _{e}^{\ast }(t)=[k_{B}T_{e}^{\ast }(t)]^{-1},
\end{equation}
%
\begin{equation}
F_{ne}(t)=-\beta _{e}^{\ast }(t)\mu _{e}^{\ast }(t),
\end{equation}
%
\begin{equation}
\mathbf{F}_{e}(t)=-\beta _{e}(t)\mathbf{v}_{e}(t),
\end{equation}
%
\begin{equation}
F_{LO}(t)=\beta _{LO}^{\ast }(t)=[k_{B}T_{LO}^{\ast }(t)]^{-1},
\end{equation}
%
\begin{equation}
F_{AC}(t)=\beta _{AC}^{\ast }(t)=[k_{B}T_{AC}^{\ast }(t)]^{-1},
\end{equation}
and we recall that $\beta _{0}$ [in Eq. (182)] is
$[k_{B}T_{0}]^{-1}$. These Eqs. (184) to (188) introduce the
so-called quasitemperatures, $T_{e}^{\ast }(t)$, $T_{LO}^{\ast
}(t)$, $T_{AC}^{\ast }(t)$, of electrons and phonons, and the
quasi-chemical potential $\mu _{e}^{\ast }(t)$ and the drift
velocity $\mathbf{v}_{e}(t)$ of the electrons; $k_{B}$ is as usual
Boltzmann constant.

Proceeding with the calculations of the equations of evolution of
the basic variables, the corresponding set of equations of evolution
are obtained, which have expressions of the form [42,59,62]
%
\begin{equation}
\frac{d}{dt}E_{e}(t)=-\frac{e}{m_{e}^{\ast }}\mathbf{E}\cdot
\mathbf{P}_{e}(t)-J_{E_{e}}^{(2)}(t)\,\,,
\end{equation}
%
\begin{equation}
\frac{d}{dt}\mathbf{P}_{e}(t)=-nVe\mathbf{E}+\mathbf{J}_{\mathbf{P}
_{e}}^{(2)}(t)+\mathbf{J}_{\mathbf{P}_{e},imp}^{(2)}(t)\,\,,
\end{equation}
%
\begin{equation}
\frac{d}{dt}E_{LO}(t)=J_{E_{LO}}^{(2)}(t)-J_{LO,AN}^{(2)}(t)\,\,,
\end{equation}
%
\begin{equation}
\frac{d}{dt}
E_{AC}(t)=J_{E_{AC}}^{(2)}(t)+J_{LO,AN}^{(2)}(t)-J_{AC,dif}^{(2)}(t)\,\,,
\end{equation}
where, we recall, $E_{e}(t)$ is the carriers' energy and
$\mathbf{P}_{e}(t)$ the linear momentum; $E_{LO}(t)$ the energy of
the \textsc{lo} phonons which strongly interact with the carriers
via Fr\"{o}hlich potential in these strong-polar semiconductors
(hence it predominates over the nonpolar-deformation potential
interaction and then the latter is disregarded); $E_{AC}(t)$ is
the energy of the acoustic phonons playing the role of a thermal
bath; and $\mathbf{E}$ stands for the constant electric field.

Let us analyze these equations term by term. In Eq. (189) the first
term on the right accounts for the rate of energy transferred from
the electric field to the carriers, and the second term accounts for
the transfer of the resulting excess energy of the carriers to the
phonons. In Eq. (190) the first term on the right is the driving
force generated by the presence of the electric field. The second
term is the rate of momentum transfer due to the interaction with
the phonons, and the last one is a result of scattering by
impurities (these two terms are then momentum relaxation
contributions). In Eq. (191) and Eq. (192) the first term on the
right describes the rate of change of the energy of the phonons due
to interaction with electrons. More precisely they account for the
gain of the energy transferred to them from the hot carriers and
then the sum of contributions $J_{E_{LO}}^{(2)}(t)$ and
$J_{E_{AC}}^{(2)}(t)$ is equal to the last in Eq. (189), but
accompanied with a change of sign. The second term in Eq. (191)
accounts for the rate of transfer of energy from the optical phonons
to the acoustic ones, via anharmonic interaction. The contribution
$J_{LO,AN}^{(2)}(t)$ is the same but with different sign in Eq.
(191) and Eq. (192). Finally, the diffusion of heat from the
\textsc{ac} phonons to the reservoir is accounted for in the last
term in Eq. (192). The detailed expression for the collision
operators are given in Ref. [42] (quantities $J$ are positive).

The solution of these equations allows for a detailed analysis of
the nonequilibrium thermodynamic state and transport properties of
these materials. Let us consider first the III-Nitride compounds,
which nowadays present a particular interest as a result of their
potential use in lasers and diodes emitting in the blue and
ultraviolet region (see for example [63]).

Let us consider the steady state which follows very rapidly (in a
hundred-fold \emph{femtosecond} time scale), what can be
understood on the basis of the action of the intense Fr\"{o}hlich
interaction in these strong polar semiconductors, with the rate of
transfer of energy from carriers to \textsc{lo} phonons rapidly
equalizing the rate of energy pumping from the external field of
intensity $\mathcal{E}$, even at high fields. It is then
characterized by the constant-in-time variables quasitemperature,
$T_{e}^{\ast }$, drift velocity, $\mathbf{v}_{e}$, and
quasi-chemical potential, $\mu _{e}^{\ast }$, all referring to the
electron system, and $T_{LO}^{\ast }$, the quasitemperature of the
\textsc{lo} phonons, and \textsc{ta}, the quasitemperature of the
acoustic phonons. In Fig. 6 it is shown the dependence of
$T_{e}^{\ast }$ with the electric field, while in Fig. 7 is
presented such dependence for the drift velocity for n-doped GaN,
with $n=$ 10$^{17}$cm$^{-3}$. The quasi-chemical potential is
determined by the values of the concentration and the electron
quasitemperature, and the deviation of $T_{LO}^{\ast }$ and
$T_{A}$ from the value in equilibrium is small and can be
neglected.

%
\begin{figure}[h]
\center
\includegraphics[width=10cm]{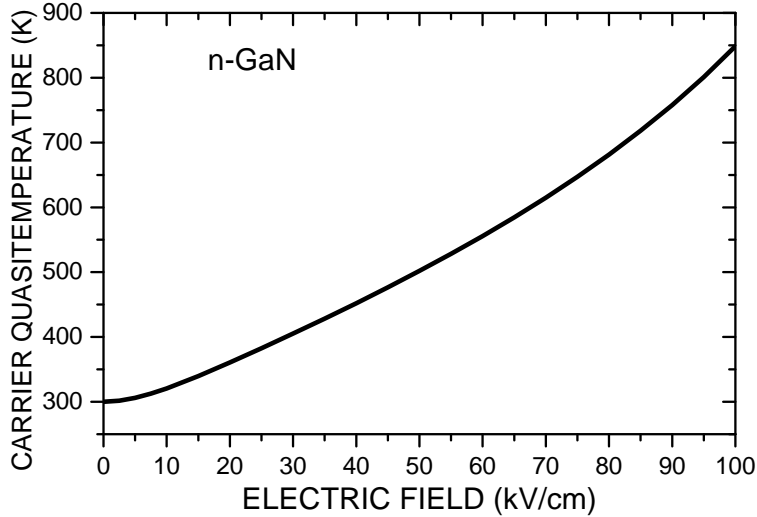}
\caption{Dependence of the quasitemperature $T_{e}^{\ast }$ with the
electric field in the steady state of n-doped GaN. We have used
$m^{\ast }=0.19m_{0}$, $\varepsilon _{0}=9.5$, a carrier
concentration $n=$ 10$^{17}$cm$^{-3}$, and the thermal bath
temperature is $T_{0}=300$ K. After Ref. [64]}
\end{figure}

We can now proceed to calculate the cross section for scattering
of light by electrons in the electric field $\mathcal{E}$. Taking
into account that the interaction of electrons and radiation is
given by
%
\begin{equation}
V_{ER}\mathbf{Q}=G(\mathbf{Q})\hat{n}_{\mathbf{Q}}+H.c.\,,
\end{equation}
where $\hbar \mathbf{Q}$ is the momentum of the photon,
$G(\mathbf{Q})$ is the matrix element of the interaction, and
%
\begin{equation}
\hat{n}_{\mathbf{Q}}=\sum\limits_{\mathbf{k}}\hat{n}_{\mathbf{kQ}
}=\sum\limits_{\mathbf{k}}c_{\mathbf{k+Q}}^{\dag
}c_{\mathbf{k}}\,,
\end{equation}
the cross section of Eq. (143), in first order in the scattering
operator, is given by
%
\begin{equation}
\frac{d^{2}\sigma (\mathbf{Q},\omega )}{d\omega d\Omega
}=\frac{V^{2}}{(2\pi \hbar )^{3}}\frac{\hbar \omega
^{2}}{c^{4}}|G(\mathbf{Q})|^{2}\int \limits_{-\infty }^{\infty
}d\tau e^{-i\omega \tau }Tr\{n_{\mathbf{Q}}^{\dag
}n_{\mathbf{Q}}\mathcal{R}_{\varepsilon }\}\,,
\end{equation}
and we recall that the system is in a steady state, and, moreover,
both integrals in Eq. (143) can be combined in the given form above,
and we have used that $g(\omega )= \hbar ^{3}\omega ^{2}/c^{3}$ and
$v_{0}=c$.

%
\begin{figure}[h]
\center
\includegraphics[width=10cm]{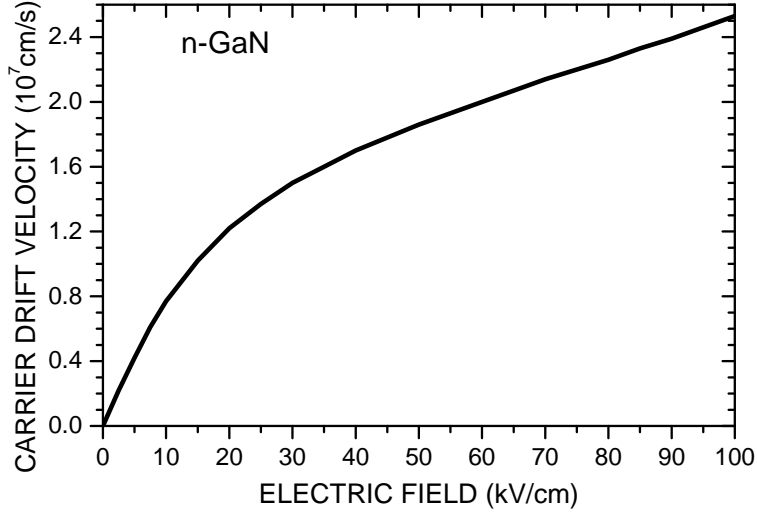}
\caption{Dependence of the drift velocity of the electrons with the
electric field in the steady state of n-doped GaN, in the same
experimental conditions indicated in the caption to Fig. 6. After
Ref. [64].}
\end{figure}

We can now use the generalized fluctuation-dissipation theorem of
the preceding section. Using the approximation of keeping only what
we have termed as the relevant part, i.e., taking
$\mathcal{\bar{R}}$ of Eq. (183) instead of
$\mathcal{R}_{\varepsilon }$, in the steady state, we find that
%
\begin{equation}
\frac{d^{2}\sigma (\mathbf{Q},\omega )}{d\omega d\Omega }\sim
\lbrack 1-e^{-\beta (\hbar \omega -\mathbf{Q}\cdot
\mathbf{v}_{e})}]^{-1} Im\, \epsilon ^{-1}(\mathbf{Q},\omega
|\mathcal{E})\,,
\end{equation}
In this Eq. (196), $\epsilon (\mathbf{Q},\omega |\mathcal{E})$ is
the dielectric function at wave vector $\mathbf{Q}$ and frequency
$\omega $ (the momentum and energy transfer in the scattering event
as we have seen), once we use that
%
\begin{equation}
\epsilon (\mathbf{Q},\omega |\mathcal{E})=1-\mathcal{V}(\mathbf{Q}
)\sum\limits_{\mathbf{k}}\frac{f(\mathbf{k+Q}|\mathcal{E})-f(\mathbf{k}|
\mathcal{E})}{\mathcal{E}(\mathbf{k+Q})-\mathcal{E}(\mathbf{k})-\hbar
(\omega +is)}\,,
\end{equation}
where in the effective mass approximation
$\mathcal{E}(\mathbf{k})=\hbar ^{2}k^{2}/2m_{e}^{\ast }$, and Eq.
(197) is of the form of Lindhardt (RPA) dielectric function, but in
terms of the nonequilibrium distribution functions
$f(\mathbf{k}|\mathcal{E})$. They have an expression of a drifted
Fermi-Dirac-like distribution (with the presence of the electric
field-dependent quasitemperature and quasi-chemical potential),
which in the usual experimental condition can be appropriately
approximated by a drifted Maxwell-Boltzmann-like distribution,
namely
%
\begin{equation}
f(\mathbf{k}|\mathcal{E})=A\exp \{-\beta
(\mathcal{E})[E(\mathbf{k})-\mathbf{E}\cdot \hbar \mathbf{k}]\}\,,
\end{equation}
with
%
\begin{equation}
\beta _{e}^{\ast }(\mathcal{E})=\frac{1}{k_{B}T^{\ast
}(\mathcal{E})}\,,
\end{equation}
%
\begin{equation}
A(\mathcal{E})=\frac{8\pi ^{3}n\hbar ^{3}[\beta _{e}^{\ast
}(\mathcal{E})]^{3/2}}{(2\pi m_{e}^{\ast })^{3/2}}\,,
\end{equation}
where it has been used that $\mathcal{V}(\mathbf{Q})=4\pi
e^{2}/(V\epsilon _{0}Q^{2})$ is the Fourier transform of the Coulomb
potential with $\epsilon _{0}$ being the static dielectric constant
and $V$ the volume of the system. Moreover, $s$ is a positive
infinitesimal which is taken in the limit of going to $+0$ to
produce the real and imaginary parts of $\epsilon (\mathbf{Q},\omega
)=\epsilon _{1}(\mathbf{Q},\omega )+i\epsilon _{2}(\mathbf{Q},\omega
)$. Going over the continuum, i.e., transforming the summation in
Eq. (197) in an integral and using spherical coordinates $k$,
$\theta $, $\varphi $ we find for the real part of the dielectric
function [65]
%
\begin{equation}
\epsilon _{1}(\mathbf{Q},\omega
)=1-\frac{\mathcal{V}(\mathbf{Q})}{4\pi ^{3}} F(\mathbf{Q},\omega
)\,,
\end{equation}
where
%
\begin{equation}
F(\mathbf{Q},\omega )=-\frac{n\pi ^{3}\sqrt{2^{5}m_{e}^{\ast
}\beta _{e}^{\ast }}}{\hbar Q}[D(y_{1})+D(y_{2})]\,,
\end{equation}
with
%
\begin{equation}
D(y)=\exp (-y^{2})\int\limits_{0}^{y}\exp (x^{2})dx\,,
\end{equation}
which is Dawson's integral, and
%
\begin{equation}
y_{1}=\sqrt{\frac{\beta _{e}^{\ast }\hbar ^{2}}{2m_{e}^{\ast
}}}\left[ \frac{ Q}{2}+\frac{m_{e}^{\ast }}{\hbar Q}(\omega
-\mathbf{Q}\cdot \mathbf{v}_{e}) \right] \,,
\end{equation}
%
\begin{equation}
y_{2}=\sqrt{\frac{\beta _{e}^{\ast }\hbar ^{2}}{2m_{e}^{\ast
}}}\left[ \frac{ Q}{2}-\frac{m_{e}^{\ast }}{\hbar Q}(\omega
-\mathbf{Q}\cdot \mathbf{v}_{e}) \right] \,,
\end{equation}

Substituting Eq. (202) in Eq. (201), we obtain that
%
\begin{equation}
\epsilon _{1}(\mathbf{Q},\omega )=1+\sqrt{\frac{2m_{e}^{\ast
}}{\beta _{e}^{\ast }}}\frac{k_{DH}^{2}}{\hbar
Q^{3}}[D(y_{1})+D(y_{2})]\,,
\end{equation}
where
%
\begin{equation}
k_{DH}^{2}=\frac{4\pi e^{2}n}{\epsilon _{0}k_{B}T_{e}^{\ast }}\,,
\end{equation}
is the Debye-Huckel screening factor. On the other hand, the
imaginary part is given by
%
\begin{equation}
\epsilon _{2}(\mathbf{Q},\omega )=\frac{\pi \sqrt{m_{e}^{\ast
}}k_{DH}^{2}}{2\beta _{e}^{\ast }\hbar Q^{3}}[\exp
(-y_{2}^{2})-\exp (-y_{1}^{2})]\,.
\end{equation}
%

%
\begin{figure}[t]
\center
\includegraphics[width=10.0cm]{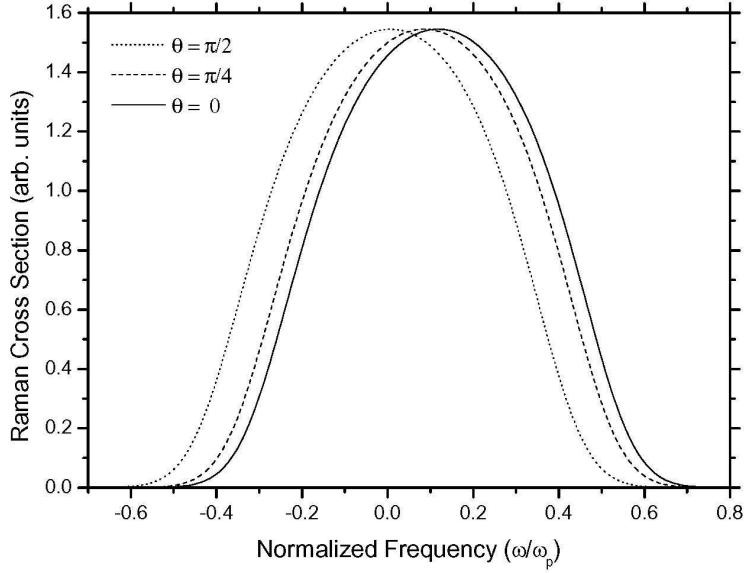}
\caption{The Raman band due to scattering by the single-particle
elementary excitations: $\mathcal{E}=$ 100 kV/cm, $Q=1.8\times
10^{5}$cm$^{-1}$, the thermodynamic state as characterized by Figs.
6 and 7 for the given $\mathcal{E}$, and the three scattering
angles, $\theta =0$ (full line), $\theta =\pi /4$ (dashed line), and
$\theta =\pi /2$ (dotted line). After Ref. [65].}
\end{figure}
%
%
\begin{figure}[t]
\center
\includegraphics[width=10.0cm]{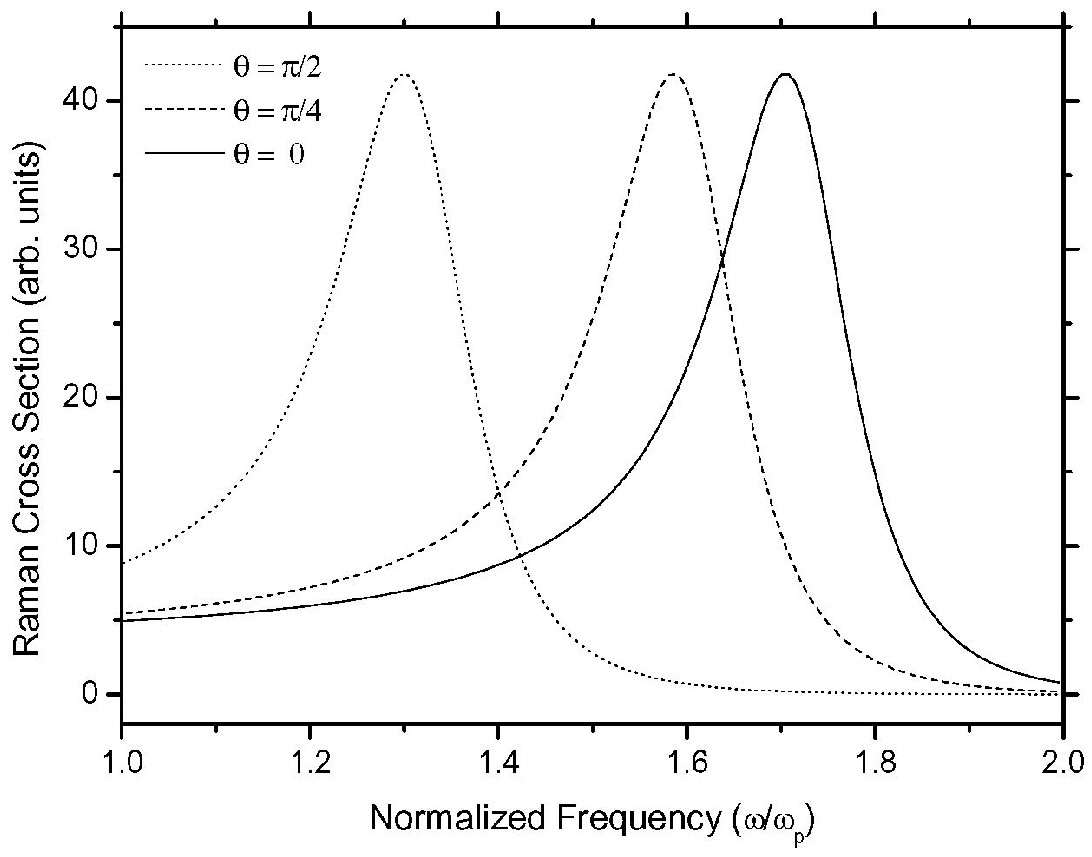}
\caption{The Raman band due to scattering by the plasma excitation:
same specifications as in the caption to Fig. 8, but for $Q=5\times
10^{4}$cm$^{-1}$. After Ref. [65].}
\end{figure}

Let us consider the case of GaN in the steady-state thermodynamic
conditions indicated in Figs. 6 and 7, in the case:
$\mathcal{E}=100$ kV/cm, and three experimental geometries, i.e.
three values of the scattering angle $\theta$. The scattering
spectrum is composed of two contributions, namely, scattering, in
Fig. 8 with $Q=1.8\times 10^{5}$cm$^{-1}$, by individual electrons
at low $\omega $, and in Fig. 9 scattering, with $Q=5\times
10^{4}$cm$^{-1}$, by collective excitations (plasmons) around the
plasma frequency $\omega^{2} _{p}=4\pi ne^{2}/\epsilon
_{0}m_{e}^{\ast }$. In these Figs. 8 and 9 it can be noticed a shift
in the scattering bands, depending on the scattering angle. This is
a result of the presence of the term $\mathbf{Q}\cdot
\mathbf{v}_{e}$ in Eq. (196), which is null for the experimental
geometry in which the momentum transfer $\mathbf{Q}$ is
perpendicular to the drift velocity, which is in the direction of
the electric field ($\mathbf{v}_{e}=\mathcal{M}_{e}\mathbf{E}$,
where $\mathcal{M}_{e}$ is the electron mobility), and maximum when
$\mathbf{Q}$ and $\mathbf{v}_{e}$ are parallel. This has a quite
interesting consequence, consisting in that the shift in frequency
permits a measurement of the drift velocity, and then of the
mobility in the conditions of the experiment. If we call $\Delta
\omega _{peak}$ the difference of frequencies at the peak positions
of the bands for scattering by plasmons, for $\theta =0$ o and
$\theta =\pi /2$ (Fig. 9), then $v_{e}=\Delta \omega _{peak}/Q$ and
$\mathcal{M}_{e}=v_{e}/\mathcal{E}=\Delta \omega
_{peak}/Q\mathcal{E}$.

\section{Final Remarks}

We have presented a Response Function Theory, accompanied of a
Fluctuation Dissipation Theorem, and a Theory of Scattering adapted
to deal with systems arbitrarily away from equilibrium, including
situations of time and space experimental resolution.

Such theory was built within the framework of a Gibbs-style
Non-Equilibrium Statistical Formalism. The general form of the
generalized susceptibility (space and time dependent) is obtained in
the form of space and time dependent correlation functions defined
over the nonequilibrum ensemble. Moreover it is dependent on the
variables that characterized the nonequilibrium thermodynamic state
of the system. Therefore, the generalized susceptibility of the
corresponding experimental situation is coupled to the equations of
evolution of the nonequilibrium thermodynamic variables. It is also
presented the method eventually useful for calculations, of
nonequilibrium thermodynamic Green functions, that is, the extension
to arbitrary nonequilibrium conditions of Bogoliubov-Tyblikov
thermodynamic Green functions.

A Fluctuation-Dissipation theorem in the context of the
Nonequilibrium Ensemble Formalism has been presented in Section III
(see also subsection IV.A), which is an extension to arbitrary
nonequilibrium conditions of Kubo's one.

In Section V we have presented a Theory of Scattering appropriate
for scattering experiments done on systems which are arbitrarily
away from equilibrium. A space and time dependent scattering cross
section is obtained, which, as in the case of the generalized
susceptibility of Response Function, is dependent on the
nonequilibrium thermodynamic state of the system, and then the
scattering cross section is coupled to the equations of evolution of
the nonequilibrium variables.

Finally, in section VI we have presented some illustrative examples
of the working of the theory in the analysis of several experimental
situations, namely, in ultrafast laser spectroscopy and charge
transport in doped polar semiconductors.

\textbf{Acknowledgments:} The authors would like to acknowledge
partial financial support received from the S\~{a}o Paulo State
Research Agency (FAPESP) and the Brazilian National Research Council
(CNPq): The authors are CNPq Research Fellows.

\newpage
\appendix

\section{The Nonequilibrium Statistical Operator}

Construction of nonequilibrium statistical ensembles, that is, a
Nonequilibrium Statistical Ensemble Formalism, NESEF for short
[8-11], consisting in, basically, the derivation of a nonequilibrium
statistical operator (probability distribution in the classical
case) has been attempted along several lines. In a brief summarized
way we descrive the construction of NESEF within a heuristic
approach, and, first, it needs to be noticed that for systems away
form equilibrium, several important points need to be carefully
taken into account in each case under consideration:

\begin{enumerate}
\item \emph{The choice of the basic variables} (a wholly different choice
than in equilibrium when it suffices to take a set of those which
are constants of motion), which is to be based on an analysis of
what sort of macroscopic measurements and processes are actually
possible, and moreover, one is to focus attention not only on what
can be observed but also on the character and expectative concerning
the equations of evolution for these variables [11,66]. We also
notice that even though at the initial stage we would need to
introduce all the observables of the system, an eventually
variances, as time elapses more and more contracted descriptions can
be used when it enters into play Bogoliubov's principle of
correlation weakening and the accompanying hierarchy of relaxation
times [67].

\item \emph{The question of irreversibility} (or Eddington's arrow of time)
on what Rudolf Peierls stated that: "In any theoretical treatment of
transport problems, it is important to realize at what point the
irreversibility has been incorporated. If it has not been
incorporated, the treatment is wrong. A description of the situation
that preserves the reversibility in time is bound to give the answer
zero or infinity for any conductivity. If we do not see clearly
where the irreversibility is introduced, we do not clearly
understand what we are doing" [68].

\item \emph{Historicity needs be introduced}, that is, the idea that it must
be incorporated all the past dynamics of the system (or historicity
effects), all along the time interval going from a starting
description of the macro-state of the sample in the given
experiment, say at $t_{0}$, up to the time $t$ when a measurement is
performed. This is a quite important point in the case of
dissipative systems as emphasized among others by John Kirkwood,
Green, Robert Zwanzig and Hazime More [15-19]. It implies in that
the history of the system is not merely the series of events in
which the system has been involved, but it is the series of
transformations along time by which the system progressively comes
into being at time $t$ (when a measurement is performed), through
the evolution governed by laws of mechanics. [69]
\end{enumerate}

Concerning the question of the choice of the basic variables,
differently to the case in equilibrium, immediately after the open
system of $N$ particles, in contact with external sources and
reservoirs, has been driven out of equilibrium, it would be
necessary to describe its state in terms of all its observables and,
eventually, introducing direct and cross-correlation. But, as time
elapses Bogoliubov's principle of correlation weakening allow us to
introduced increasing contractions of descriptions. Let us say that
we can introduce a description based on the observables
$\{\hat{P}_{j}\}$, $j=1,2,...,n$, on which depends the noneuilibrium
statistical operator.

On the question of irreversibility Nicolai S. Krylov [21] considered
that there always exists a physical interaction between the measured
system and the external world that is constantly "jolting" the
system out of its exact microstate. Thus, the instability of
trajectories and the unavoidable finite interaction with the outside
would guarantee the working of a "crudely prepared" macroscopic
description. In the absence of a proper way to introduce such
effect, one needs to resort to the \emph{interventionist's
approach}, which is grounded on the basis of such ineluctable
process of randomization leading to the asymmetric evolution of the
macro-state.

The "intervention" consists into introducing in the Liouville
equation of the statistical operator, of the otherwise isolated
system, a particular source accounting for Krylov's "\emph{jolting}"
\emph{effect}, in the form (written for the logarithm of the
statistical operator)
%
\begin{equation}
\frac{\partial }{\partial t}\ln \Re _{\varepsilon
}(t)+\frac{1}{i\hbar }[\ln \Re _{\varepsilon
}(t),\hat{H}]=-\varepsilon \lbrack \ln \Re _{\varepsilon }(t)-\ln
\bar{\Re}(t,0)],
\end{equation}
where $\varepsilon $ (kind of reciprocal of a relaxation time) is
taken to go to $+0$ after the calculations of average values has
been performed. Such mathematically inhomogeneous term, in the
otherwise normal Liouville equation, implies in a continuous
tendency of relaxation of the statistical operators towards a
\emph{referential distribution}, $\bar{\Re}$, which, as discussed
below, represents an instantaneous quasi-equilibrium condition.

We can see that Eq. (A.1) consists of a regular Liouville equation
but with an infinitesimal source, which provides Bogoliubov's
symmetry breaking of time reversal and is responsible for
disregarding the advanced solutions [8,11,70]. This is described by
a Poisson distribution and the result at time $t$ is obtained by
averaging over all $t^{\prime }$ in the interval $(t_{0},t)$, once
the solution of Eq. (A.1) is
%
\begin{equation}
\Re _{\varepsilon }(t)=\exp \left\{ -\hat{S}(t,0)+\int
\limits_{t_{0}}^{t}dt^{\prime }e^{\varepsilon (t^{\prime
}-t)}\frac{d}{dt^{\prime }}\hat{S}(t^{\prime },t^{\prime
}-t)\right\} ,
\end{equation}
where
%
\begin{equation}
\hat{S}(t,0)=-\ln \bar{\Re}(t,0),
\end{equation}
%

%
\begin{equation}
\hat{S}(t^{\prime },t^{\prime }-t)=\exp \left\{ -\frac{1}{i\hbar
}(t^{\prime }-t)\hat{H}\right\} \hat{S}(t^{\prime },0)\exp \left\{
\frac{1}{i\hbar }(t^{\prime }-t)\hat{H}\right\} ,
\end{equation}
and the initial-time condition at time $t_{0}$, when the formalism
begins to be applied, is
%
\begin{equation}
\Re _{\varepsilon }(t_{0})=\bar{\Re}(t_{0},0).
\end{equation}

In $\bar{\Re}$ and $\hat{S}$, the first time variable in the
argument refers to the evolution of the nonequilibrium thermodynamic
variables and the second to the time evolution of the dynamical
variables, both of which have an effect on the operator.

This time $t_{0}$, of initiation of the statistical description, is
usually taken in the remote past ($t_{0}\rightarrow -\infty $)
introducing an adiabatic switching-on of the relaxation process, and
in Eq. (A.2) the integration in time in the interval $(t_{0},t)$ is
weighted by the kernel $\exp \{ \varepsilon (t^{\prime }-t) \}$. The
presence of this kernel introduces a kind of \emph{evanescent
history} as the system macro-state evolves toward the future from
the boundary condition of Eq. (A.5) at time ($t_{0}\rightarrow
-\infty $) a fact evidenced in the resulting kinetic theory
[8-11,14,17] which clearly indicates that it has been introduced a
\emph{fading memory} of the dynamical process. It can be noticed
that the statistical operator can be write in the form
%
\begin{equation}
\Re _{\varepsilon }(t)=\bar{\Re}(t,0)+\Re^{\prime } _{\varepsilon
}(t)\;.
\end{equation}
involving the auxiliary probability distribution $\bar{\Re}(t,0)$,
plus $\Re^{\prime } _{\varepsilon }(t)$ which contains the
historicity and irreversibility effects. Moreover, in most cases we
can consider the system as composed of the system of interest (on
which we are performing an experiment) in contact with ideal
reservoirs. Thus, we can write
%
\begin{equation}
\bar{\Re}(t,0) = \bar{\rho}(t,0)\times \rho_{R} \;.
\end{equation}
and
%
\begin{equation}
\Re _{\varepsilon }(t)=\rho _{\varepsilon }(t)\times \rho _{R} \;,
\end{equation}
where $\rho _{\varepsilon }(t)$ is the statistical operator of the
nonequilibrium system, $\bar{\rho}$  the auxiliary one, and $\rho
_{R}$ the stationary one of the ideal reservoirs, with $\rho
_{\varepsilon }(t)$ given then by
%
\begin{equation}
\rho _{\varepsilon }(t)=\exp \left\{
-\hat{S}(t,0)+\int\limits_{-\infty }^{t}dt^{\prime }e^{\varepsilon
(t^{\prime }-t)}\frac{d}{dt^{\prime }}\hat{S}(t^{\prime },t^{\prime
}-t)\right\} \;,
\end{equation}
having the initial value $\bar{\rho}(t_{0},0)$ ($t_{0}\rightarrow
-\infty $), and where
%
\begin{equation}
\hat{S}(t,0)=-\ln \bar{\rho}(t,0) \;,
\end{equation}

Finally, it needs be provided the auxiliary statistical operator
$\bar{\rho}(t,0)$. It defines an instantaneous distribution at time
$t$, which describes a "frozen" equilibrium defining at such given
time the macroscopic state of the system, and for that reason is
sometimes dubbed as the \emph{quasi-equilibrium statistical
operator}. On the basis of this (or, alternatively, via the extremum
principle procedure [11,71-76], and considering the description of
the non-equilibrium state of the system in terms of the basic set of
dynamical variables $\hat{P}_{j}$, the reference or instantaneous
quasi-equilibrium statistical operator is taken as a canonical-like
one given by
%
\begin{equation}
\bar{\rho}(t,0)=\exp \{-\phi
(t)-\sum\limits_{j}^{n}F_{j}(t)\hat{P}_{j}\}\;,
\end{equation}
in the classical case, with $\phi (t)$ ensuring the normalization of
$\bar{\rho}$ and playing the role of a kind of a logarithm of a
partition function, say, $\phi (t)=\ln \bar{Z}(t)$. Moreover, in
this Eq. (A.11), $F_{j}$ are the nonequilibrium thermodynamic
variables associated to each kind of basic dynamical variables
$\hat{P}_{j}$. The nonequilibrium thermodynamic space of states is
composed by the basic variables $\{ Q_{j}(t) \}$ consisting of the
averages of the $\{ \hat{P}_{j} \}$ over the nonequilibrium
ensemble, namely
%
\begin{equation}
Q_{j}(t)=Tr\{\hat{P}_{j}\rho _{\varepsilon }(t)\}\,,
\end{equation}
which are then functionals of the $\{F_{j}(t)\}$ and there follow
the equations of state
%
\begin{equation}
Q_{j}(t)=-\frac{\delta \phi (t)}{\delta F_{j}(t)}=-\frac{\delta \ln
\bar{Z}(t)}{\delta F_{j}(t)}\,,
\end{equation}
where $\delta $ stands for functional derivative.

Moreover
%
\begin{equation}
\bar{S}(t)=Tr\{
\hat{\bar{S}}(t,0)\bar{\rho}(t,0)\}=-Tr\{\bar{\rho}(t,0)\ln
\bar{\rho}(t,0)\}\,,
\end{equation}
is the so-called informational entropy characteristic of the
distribution $\bar{\rho}$, a functional of the basic variables
$\{Q_{j}(t)\}$, and it is verified the alternative form of the
equations of state given by
%
\begin{equation}
-\frac{\delta \bar{S}(t)}{\delta Q_{j}(t)}=F_{j}(t)\,,
\end{equation}

\end{document}